%% file: sn_hmm.tex
\documentclass[12pt,a4paper]{article}
\usepackage{amssymb}
\usepackage{amsmath}
\usepackage{cite}
\usepackage{a4p}
\usepackage{graphics}
\usepackage{epsfig}
\usepackage{rotating}
\usepackage{atlas}
\usepackage{Lep}
\usepackage{marnote}
\usepackage{physics}
\usepackage{captions_my}
\usepackage{xspace}
\usepackage{ifthen,Lep}
\begin{document}
\bibliographystyle{l3style}
\input{higgs_atlas.tex}
\input{Abstract}

\clearpage
\input{Introduction}

\input{MSSM}
\input{Exp}

\input{Simul}

\clearpage
\bibliography{sn_hmm}
\end{document}

%% file: higgs_atlas.tex
\newcommand{\MR}{\mathrm}

\newcommand{\red}{\color{red}}
\newcommand{\green}{\color{green}}
\newcommand{\blue}{\color{blue}}
\newcommand{\cyan}{\color{cyan}}
\newcommand{\black}{\color{black}}
\newcommand{\magenta}{\color{magenta}}
\newcommand{\yellow}{\color{yellow}}

\def\MZ{\ensuremath{m_{\mathrm{Z}}}}%
\def\MW{\ensuremath{m_{\mathrm{W}}}}%
\def\Mt{\ensuremath{m_{\mathrm{t}}}}%
\def\Mb{\ensuremath{m_{\mathrm{b}}}}%
\def\MH{\ensuremath{m_{\mathrm{H}}}}%
\def\Mtau{\ensuremath{m_{\tau}}}%
\def\Mmu{\ensuremath{m_{\mu}}}%
\def\Me{\ensuremath{m_{e}}}%
\def\sw{\ensuremath{\sin\vartheta_{\mathrm{W}}}}%
\def\cw{\ensuremath{\cos\vartheta_{\mathrm{W}}}}%
\def\swsq{\ensuremath{\sin^2\!\vartheta_{\mathrm{W}}}}%
\def\cwsq{\ensuremath{\cos^2\!\vartheta_{\mathrm{W}}}}%
\def\swsqb{\ensuremath{\sin^2\!\bar{\vartheta}_{\mathrm{W}}}}%
\def\cwsqb{\ensuremath{\cos^2\!\bar{\vartheta}_{\mathrm{W}}}}%
\def\swsqon{\ensuremath{\swsq\equiv 1-\MW^2/\MZ^2}}%
\def\gL{\ensuremath{g_{\mathrm{L}}}}
\def\gR{\ensuremath{g_{\mathrm{R}}}}
\def\gV{\ensuremath{g_{\mathrm{V}}}}
\def\gA{\ensuremath{g_{\mathrm{A}}}}
\def\gVb{\ensuremath{\bar{g}_{\mathrm{V}}}}
\def\gAb{\ensuremath{\bar{g}_{\mathrm{A}}}}
\def\rhob{\ensuremath{\bar{\rho}}}
%
\def\Photon{\ensuremath{\mathrm {\gamma}}}
\def\Ho{\ensuremath{\mathrm {H^0}}}
\def\Zo{\ensuremath{\mathrm {Z}}}
\def\Wp{\ensuremath{\mathrm {W^+}}}
\def\Wm{\ensuremath{\mathrm {W^-}}}
\def\Wpm{\ensuremath{\mathrm {W^\pm }}}
\def\GW{\ensuremath{\Gamma_{\mathrm{W}}}}
\def\GZ{\ensuremath{\Gamma_{\mathrm{Z}}}}
\def\GE{\ensuremath{\Gamma_{\mathrm{e}}}}
\def\GM{\ensuremath{\Gamma_{\mathrm\mu }}}
\def\GT{\ensuremath{\Gamma_{\mathrm{\tau}}}}
\def\Ghad{\ensuremath{\Gamma_{\mathrm{had}}}}
\def\GL{\ensuremath{\Gamma_{\ell}}}
\def\GN{\ensuremath{\Gamma_{\mathrm{\nu}}}}
\def\Gbb{\ensuremath{\Gamma_{\mathrm{b}}}}
\def\Gcc{\ensuremath{\Gamma_{\mathrm{c}}}}
\def\Gf{\ensuremath{\Gamma_{\mathrm{f}}}}
\def\Ginv{\ensuremath{\Gamma_{\mathrm{inv}}}}       
\def\Nnu{\ensuremath{N_{\mathrm{\nu}}}}
%
%
\def\EE{\ensuremath{\mathrm{e^+ e^-}}}%
\def\MM{\ensuremath{\mathrm{\mu ^+ \mu ^-}}}%
\def\TT{\ensuremath{\mathrm{\tau^+ \tau^-}}}%
\def\FF{\ensuremath{\mathrm{f\/ \bar{f}}}}%
\def\NN{\ensuremath{\mathrm{\nu\bar{\nu}}}}%
\def\LL{\ensuremath{\ell^+ \ell^-}}%
\def\WW{\ensuremath{\mathrm{W^+W^-}}}%
\def\ZZ{\ensuremath{\mathrm{ZZ}}}%
\def\pp{\ensuremath{\mathrm{p\bar{p}}}}%
\def\qq{\ensuremath{\mathrm{q\bar{q}}}}%
\def\dd{\ensuremath{\mathrm{d\bar{d}}}}%
\def\uu{\ensuremath{\mathrm{u\bar{u}}}}%
\def\cc{\ensuremath{\mathrm{c\bar{c}}}}%
\def\ss{\ensuremath{\mathrm{s\bar{s}}}}%
\def\bb{\ensuremath{\mathrm{b\bar{b}}}}%
\def\tt{\ensuremath{\mathrm{t\bar{t}}}}%
\def\gg{\ensuremath{\mathrm{gg}}}%
\def\Zbb{\ensuremath{\Zo \rightarrow \bb}}
\def\Zqq{\ensuremath{\Zo \rightarrow \qq}}
\def\EEHAD{\ensuremath{\EE \rightarrow \mathrm{hadrons}}}
\def\EEEE{\ensuremath{\EE \rightarrow \EE}}
\def\EEMM{\ensuremath{\EE \rightarrow \MM}}
\def\EETT{\ensuremath{\EE \rightarrow \TT}}
\def\EEFF{\ensuremath{\EE \rightarrow \FF}}
\def\EELL{\ensuremath{\EE \rightarrow \LL}}
\def\EEBB{\ensuremath{\EE \rightarrow \bb}}
\def\EECC{\ensuremath{\EE \rightarrow \cc}}
\def\EENNG{\ensuremath{\EE \rightarrow \NN \gamma}}
\def\EEWW{\ensuremath{\EE \rightarrow \WW}}

\def\mm{\ensuremath{\mathrm{\mu ^+ \mu ^-}}}%
\def\tautau{\ensuremath{\mathrm{\tau^+ \tau^-}}}%
\def\Photon{\ensuremath{\mathrm {\gamma}}}
\def\EEHADG{\ensuremath{\EE \rightarrow \mathrm{hadrons}(\Photon)}}
\def\EEEEG{\ensuremath{\EE \rightarrow \EE(\Photon)}}
\def\EEMMG{\ensuremath{\EE \rightarrow \mm(\Photon)}}
\def\EETTG{\ensuremath{\EE \rightarrow \tautau(\Photon)}}
\def\EELLG{\ensuremath{\EE \rightarrow \ell^+\ell^-(\Photon)}}
\def\ff{\ensuremath{\mathrm{f\bar{f}}}}
\def\EEFFG{\ensuremath{\EE \rightarrow \ff(\Photon)}}
\def\EENNG{\ensuremath{\EE \rightarrow \nu\bar{\nu}\Photon(\Photon)}}
\def\EEGGG{\ensuremath{\EE \rightarrow \Photon\Photon(\Photon)}}
\def\EEWWG{\ensuremath{\EE \rightarrow \MR{W^+W^-(\Photon)}}}
\def\EELNLNG {\ensuremath{\EE \rightarrow \mathrm{\ell\nu\ell\nu(\gamma)}}}
\def\EEQQENG {\ensuremath{\EE \rightarrow \mathrm{qq e\nu(\gamma)}}}
\def\EEQQMNG {\ensuremath{\EE \rightarrow \mathrm{qq \mu \nu(\gamma)}}}
\def\EEQQTNG {\ensuremath{\EE \rightarrow \mathrm{qq \tau\nu(\gamma)}}}
\def\EEQQLNG {\ensuremath{\EE \rightarrow \mathrm{qq \ell\nu(\gamma)}}}
\def\EEQQQQG {\ensuremath{\EE \rightarrow \mathrm{qq qq     (\gamma)}}}

%
%
\def\TeV{\ensuremath{\mathrm{Te\kern -0.1em V}}}
\def\GeV{\ensuremath{\mathrm{Ge\kern -0.1em V}}}
\def\MeV{\ensuremath{\mathrm{Me\kern -0.1em V}}}
\def\keV{\ensuremath{\mathrm{ke\kern -0.1em V}}}
\def\eV{\ensuremath{\mathrm{e\kern -0.1em V}}}

\def\fb{\ensuremath{\mathrm{fb}}}
\def\pb{\ensuremath{\mathrm{pb}}}
\def\nb{\ensuremath{\mathrm{nb}}}
\def\fbinv{\ensuremath{\mathrm{fb^{-1}}}}
\def\pbinv{\ensuremath{\mathrm{pb^{-1}}}}
\def\nbinv{\ensuremath{\mathrm{nb^{-1}}}}
\def\cm2{\ensuremath{\mathrm{cm^{-2}}}}
\def\sinv{\ensuremath{\mathrm{s^{-1}}}}

\def\L{\ensuremath{\cal L}}
\def\IL{\ensuremath{\int\!{\cal L}\, \mathrm{d}t}}
\def\PT{\ensuremath{P_{\mathrm{T}}}}
\def\AFB{\ensuremath{A_{\mathrm{FB}}}}
\def\AFBZ{\ensuremath{A^0_{\mathrm{FB}}}}
\def\ALR{\ensuremath{A_{\mathrm{LR}}}}
\def\RS{\ensuremath{\sqrt{s}}}
\def\RSprime{\ensuremath{\sqrt{s^\prime}}}
\def\as{\ensuremath{\alpha_{\mathrm{s}}}}
\def\asMZ{\ensuremath{\alpha_{\mathrm{s}}(\MZ)}}
\def\aqed{\ensuremath{\alpha}}
\def\Gmu{\ensuremath{G_{\mathrm{F}}}}
\def\Ebeam{\ensuremath{E_{\mathrm{b}}}}
\def\Evis{\ensuremath{E_{\mathrm{vis}}}}

\def\s0had{\ensuremath{\sigma^0_{\mathrm{had}}}}
\def\RE{\ensuremath{R_{\mathrm{e}}}}
\def\RM{\ensuremath{R_{\mathrm\mu }}}
\def\RT{\ensuremath{R_{\mathrm{\tau}}}}
\def\RL{\ensuremath{R_{\mathrm{\ell}}}}
\def\AFBZe{\ensuremath{\AFB^{\mathrm{0,e}}}}
\def\AFBZm{\ensuremath{\AFB^{\mathrm{0,\mu }}}}
\def\AFBZt{\ensuremath{\AFB^{\mathrm{0,\tau}}}}
\def\AFBZl{\ensuremath{\AFB^{\mathrm{0,\ell}}}}
\def\AFBZb{\ensuremath{\AFB^{0,\mathrm{b}}}}
\def\AFBZc{\ensuremath{\AFB^{0,\mathrm{c}}}}
\def\AFBZs{\ensuremath{\AFB^{0,\mathrm{s}}}}
\def\AFBZq{\ensuremath{\AFB^{0,\mathrm{q}}}}
\def\gVbe{\ensuremath{\gVb^{\mathrm{e}}}}
\def\gAbe{\ensuremath{\gAb^{\mathrm{e}}}}
\def\gVbm{\ensuremath{\gVb^{\mathrm\mu }}}
\def\gAbm{\ensuremath{\gAb^{\mathrm\mu }}}
\def\gVbt{\ensuremath{\gVb^{\mathrm{\tau}}}}
\def\gAbt{\ensuremath{\gAb^{\mathrm{\tau}}}}
\def\gVbl{\ensuremath{\gVb^{\mathrm{\ell}}}}
\def\gAbl{\ensuremath{\gAb^{\mathrm{\ell}}}}
\def\gVbf{\ensuremath{\gVb^{\mathrm{f}}}}
\def\gAbf{\ensuremath{\gAb^{\mathrm{f}}}}
\def\swsq{\ensuremath{\sin^2\!\theta_{\mathrm{W}}}}%
\def\cwsq{\ensuremath{\cos^2\!\theta_{\mathrm{W}}}}%
\def\ctwsq{\ensuremath{\cot^2\!\theta_{\mathrm{W}}}}%
\def\swsqb{\ensuremath{\sin^2\!\overline{\theta}_{\mathrm{W}}}}%
\def\cwsqb{\ensuremath{\cos^2\!\overline{\theta}_{\mathrm{W}}}}%

\def\susy#1{\ensuremath{\tilde{\mathrm{#1}}}}%
\def\slepton   #1{\ensuremath{\susy{\ell}^{#1}}}
\def\selectron #1{\ensuremath{\susy{\MR{e}}^{#1}}}
\def\smuon     #1{\ensuremath{\susy\mu ^{#1}}}
\def\stau      #1{\ensuremath{\susy{\tau}^{#1}}}
\def\sneutrino {\susy{\nu}}
\def\squark    {\susy{q}}
\def\stop      #1{\ensuremath{\susy{\MR{t}}_{#1}}}
\def\sbottom   #1{\ensuremath{\susy{\MR{b}}_{#1}}}
\def\stopbar   #1{\ensuremath{\susy{\MR{\bar{t}}}_{#1}}}
\def\sbottombar#1{\ensuremath{\susy{\MR{\bar{b}}}_{#1}}}
\def\chargino  #1{\ensuremath{\susy{\chi}_1^{#1}}}
\def\neutralino#1{\ensuremath{\susy{\chi}_{#1}^0}}
\def\gravitino {\susy{\MR{G}}}
\def\gluino {\susy{g}}
\def\gapprox{\ensuremath{\stackrel{>}{\scriptstyle \sim}}}
\def\lapprox{\ensuremath{\stackrel{<}{\scriptstyle \sim}}}
\newcommand{\SWW}{\ensuremath{\sigma_{\mathrm{WW}}}}
\newcommand{\pz}{\ensuremath{\phantom{0}}}
\newcommand{\pzz}{\ensuremath{\phantom{00}}}
\newcommand{\PW}{\ensuremath{\mathrm{W}}}
\newcommand{\QQ}{\ensuremath{\mathrm{q}\bar \mathrm{q}}}
\newcommand{\EN}{\ensuremath{\mathrm{e}\nu}}
\newcommand{\MN}{\ensuremath{\mu \nu}}
\newcommand{\TN}{\ensuremath{\tau\nu}}
\newcommand{\LN}{\ensuremath{\ell\nu}}
\newcommand{\WQQ}{\ensuremath{\PW\rightarrow \mathrm{qq}}}
\newcommand{\WFF}{\ensuremath{\PW\rightarrow \mathrm{f}\,\mathrm{f}}}
\newcommand{\WEN}{\ensuremath{\PW\rightarrow \EN}}
\newcommand{\WMN}{\ensuremath{\PW\rightarrow \MN}}
\newcommand{\WTN}{\ensuremath{\PW\rightarrow \TN}}
\newcommand{\WLN}{\ensuremath{\PW\rightarrow \LN}}
\newcommand{\QQEN}{\ensuremath{\mathrm{qq}\mathrm{e\nu}}}
\newcommand{\QQMN}{\ensuremath{\mathrm{qq}\mu \nu}}
\newcommand{\QQTN}{\ensuremath{\mathrm{qq}\tau\nu}}
\newcommand{\QQLN}{\ensuremath{\mathrm{qq}\ell\nu}}
\newcommand{\QQQQ}{\ensuremath{\mathrm{qqqq}}}
\newcommand{\LNLN}{\ensuremath{\ell\nu\ell\nu}}
\newcommand{\QQENG}{\QQEN\ensuremath{(\gamma)}}
\newcommand{\QQMNG}{\QQMN\ensuremath{(\gamma)}}
\newcommand{\QQTNG}{\QQTN\ensuremath{(\gamma)}}
\newcommand{\QQLNG}{\QQLN\ensuremath{(\gamma)}}
\newcommand{\LNLNG}{\LNLN\ensuremath{(\gamma)}}
\newcommand{\QQQQG}{\QQQQ\ensuremath{(\gamma)}}
\newcommand{\dm}{\ensuremath{\Delta M}}
\def\Hpm{\ensuremath{\mathrm {H^\pm }}}
\def\GA{\ensuremath{\Gamma_{\mathrm{A}}}}
\def\Gh{\ensuremath{\Gamma_{\mathrm{h}}}}
\def\GH{\ensuremath{\Gamma_{\mathrm{H}}}}
\def\h{\ensuremath{\mathrm{h}}}
\def\Aa{\ensuremath{\mathrm{A}}}
\def\H{\ensuremath{\mathrm{H}}}
\def\v12{\ensuremath{\mathrm {\frac{ v_1}{v_2}}}}
\def\tanb{\ensuremath{\mathrm {tan \beta}}}
\def\MH{\ensuremath{m_{\mathrm{H}}}}%
\def\MA{\ensuremath{m_{\mathrm{A}}}}%
\def\Mh{\ensuremath{m_{\mathrm{h}}}}%
\def\SA{\ensuremath{\sigma_{\mathrm{A}}}}%
\def\SH{\ensuremath{\sigma_{\mathrm{H}}}}%
\def\Sh{\ensuremath{\sigma_{\mathrm{h}}}}%
\def\MHPM{\ensuremath{m_{\mathrm{H_\pm}}}}%
\def\Mg{\ensuremath{m_{\gluino}}}
\def\MSUSY{\ensuremath{M_{\mathrm{SUSY}}}}%
\def\M0{\ensuremath{m_{\mathrm{0}}}}%
\def\Mdue{\ensuremath{M_{\mathrm{2}}}}%
\def\A0{\ensuremath{A_{\mathrm{0}}}}%
\def\SMeV{\ensuremath{10^{\mathrm{3}}\MeV}}
\def\eg{{\it e.g.}}
\def\DR{\ensuremath{\Delta{\mathrm{R}}}}
\def\brem{{\it bremsstrahlung}}
\def\gbrem{\ensuremath{\gamma_{\mathrm{brem}}}}
\def\Nmu{\ensuremath{N_{\mathrm{\mu}}}}
\def\PTmu{\ensuremath{P_{\mathrm{T\mu}}}}
\def\Etamu{\ensuremath{\eta_{\mathrm{\mu}}}}
\def\Njet{\ensuremath{N_{\mathrm{jet}}}}
\def\Nbjet{\ensuremath{N_{\mathrm{bjet}}}}
\def\PTjet{\ensuremath{P_{\mathrm{Tjet}}}}
\def\Etajet{\ensuremath{\eta_{\mathrm{jet}}}}
\def\wb{\ensuremath{\mathrm{w_{btag}}}}
\def\Emiss{\ensuremath{E^{\mathrm{miss}}}}
\def\ETmiss{\ensuremath{E^{\mathrm{miss}}_\mathrm{T}}}
\def\ET{\ensuremath{E_{\mathrm{T}}}}
\def\PTmumost{\ensuremath{P_{\mathrm{T\mu 1}}}}
\def\PTmuleast{\ensuremath{P_{\mathrm{T\mu 2}}}}
\def\PTjetmost{\ensuremath{P_{\mathrm{Tjet 1}}}}
\def\PTjetleast{\ensuremath{P_{\mathrm{Tjet 2}}}}
\def\Minv{\ensuremath{M^{\mathrm{inv}}_{\mu\mu}}}
\def\SPT{\ensuremath{\Sigma |P_\mathrm{T tracks}|}}
\def\SIGN{\ensuremath{\frac{\ensuremath{S}}{\ensuremath{\sqrt{B}}}}}
\def\BBA {\ensuremath{\bb \Aa \rightarrow \mathrm{\bb\mm}}}
\def\BBh {\ensuremath{\bb \h \rightarrow \mathrm{\bb\mm}}}
\def\BBH {\ensuremath{\bb \H \rightarrow \mathrm{\bb\mm}}}
\def\BBZ {\ensuremath{\bb \Zo \rightarrow \mathrm{\bb\mm}}}
\def\ZMM {\ensuremath{\Zo \rightarrow \mathrm{\bb\mm}}}
\def\BBTT {\ensuremath{\tt  \rightarrow \mathrm{\bb\mm\NN}}}
\def\ZZ {\ensuremath{ \Zo\Zo \rightarrow \mathrm{\bb\mm}}}
\def\1cut{\ensuremath{ \Nmu~ \ge~2,~ \PTmu~ \ge~ 10 ~\GeV ,~|\Etamu| ~< 2.5 }}
\def\2cut{\ensuremath{ \Njet~ \ge~2,~ \PTjet~ \ge~ 10 ~\GeV ,~|\Etajet |~< 2.5 }}
\def\3cut{\ensuremath{ \Nbjet~ \ge~1, \PTjet~ \ge~ 15 ~\GeV,~\wb ~> 1 }}
\def\8cut{\ensuremath{ \ETmiss ~< 45~ \GeV }}
\def\4cut{\ensuremath{25~ \GeV~ <~ \PTmumost~ <~ 95~ \GeV }}
\def\5cut{\ensuremath{20~ \GeV~ <~ \PTmuleast~ <~ 60~ \GeV }}
\def\7cut{\ensuremath{  \PTjetleast ~< 40~ \GeV }}
\def\6cut{\ensuremath{ \PTjetmost~ < 70 ~\GeV }}
\def\9cut{\ensuremath{ |\Minv - \MA |~\mathrm {or}~|\Minv - \Mh| < 2\sigma }}
\def\cutdieci{\ensuremath{ \SPT ~ <~ 5 ~\GeV ~\mathrm{in~cone }~\DR ~<~0.2   }}
%
\def\StrechA{\ensuremath{S^{\mathrm{300}}_{\h,~\Aa}}}
\def\StrenhA{\ensuremath{S^{\mathrm{30}}_{\h,~\Aa}}}
\def\SdiehA{\ensuremath{S^{\mathrm{10}}_{\h,~\Aa}}}
\def\StrecH{\ensuremath{S^{\mathrm{300}}_{\H}}}
\def\StrenH{\ensuremath{S^{\mathrm{30}}_{\H}}}
\def\SdieH{\ensuremath{S^{\mathrm{10}}_{\H}}}
\def\StrechAH{\ensuremath{S^{\mathrm{300}}_{\h,~\Aa,~\H}}}
\def\StrenhAH{\ensuremath{S^{\mathrm{30}}_{\h,~\Aa,~\H}}}
\def\SdiehAH{\ensuremath{S^{\mathrm{10}}_{\h,~\Aa,~\H}}}

%% file: Abstract.tex
\begin{titlepage}
\title{The ATLAS discovery potential for \\
           MSSM neutral Higgs bosons 
       decaying  to a $\boldsymbol {\MM} $ pair\\ 
          in the mass range up to 130 GeV }

\author{ \bf Simonetta Gentile$^1$ \\
\bf Halina Bilokon$^2$, \bf Vitaliano Chiarella$^2$, \bf Giovanni Nicoletti$^2$\\
\it  (1) Dipartimento di Fisica,  Universit\`a La Sapienza, Sez. I.N.F.N., Roma,\\
\it  (2) Laboratori Nazionali di Frascati, I.N.F.N., Frascati, Roma.
 }

\begin{abstract}
Results are presented on the discovery potential for MSSM  neutral Higgs bosons in the $\Mh-{max}$ scenario. The region of  
large \tanb, between  15 and 50, and  mass between  $\approx$ 95 and  130 \GeV\  is considered in the framework of the ATLAS 
experiment at the Large Hadron Collider (LHC), for a centre-of-mass energy  \RS\ = 14 TeV. 
This parameter region is not fully covered by the present data either from LEP or from Tevatron. 

The h/A bosons, supposed to be very close in mass in that region, are studied in the channel h/A $\ra \MM$\ accompanied by 
two b-jets. The study includes a method to control the most copious background, $\Zo \ra \MM$ accompanied by two b-jets.
A possible contribution of the H boson to the signal is also considered.
\end{abstract}
%
%
\end{titlepage}

\normalsize

%% file: Introduction.tex
\section{Introduction}\label{sec:intr}

The Minimal Supersymmetric Standard Model (MSSM)  is the most investigated extension of the Standard Model (SM).

The theory requires two Higgs doublets giving origin to five Higgs bosons:
two CP-even neutral scalars, h and H (h is the lighter of the two), one CP-odd neutral scalar, A, and one pair of charged Higgs 
bosons, \Hpm~  
\cite{mssm2,Haber:1984rc,mssm3}.
The discovery of any one of these particles is a crucial element for the confirmation of the model. 
This is  a key point in the physics program of future accelerators and in particular of the LHC.

After the conclusion of the LEP program in the year 2000, the experimental limit on the mass of the Standard Model Higgs 
boson \H\  was established at 114.4 GeV with 95\% CL \cite{Barate:2003sz}. Limits were also set on the mass of neutral 
\cite{Schael:2006as} and charged \cite{unknown:2001xy} MSSM Higgs bosons for most of the representative sets of model parameters.
\par 
The motivation of this study is to explore the potential of the ATLAS detector for the discovery of neutral MSSM Higgs bosons 
in the parameter region not excluded by the LEP and Tevatron data. We shall focus on the search for \h, the lightest of the 
neutral Higgs bosons. Its mass, taking account of  radiative corrections, is predicted to be smaller than 140 \GeV, see 
\cite{Schael:2006as} 
and references therein. The search for a mass close to the mass of the Z boson,~\MZ,  will be a challenging test of detector 
performance and of the analysis method of disentangling the signal from the background.
\par 
In the first part of this paper we review the MSSM framework, the production mechanism in hadron collisions, the present 
experimental situation and the discovery potential at the LHC. 
\par
In the second part, after describing the Monte Carlo generator and the software tools used, we discuss the detector performance  
relevant for this search of h, A and H, the analysis strategy and the results of the scan over the MSSM (\MA, \tanb ) plane. 
Details of this analysis are given in  Ref. \cite{Gentile07} and Ref. \cite{Gentile06}. 
\par 
In the conclusion, results are presented on the neutral MSSM  Higgs bosons discovery potential at the LHC based on the ATLAS 
detector.

%% file: MSSM.tex
 \section{Minimal Supersymmetric Standard Model}\label{sec:MSSM}

We discuss a few points of the model, useful for the present analysis. For a  complete 
review see Refs. \cite{Martin:1997ns,Haber:1984rc}.
At tree level, the masses of the five Higgs bosons of the MSSM are related by the following equations:  

\begin{equation}
m_{H,h}^2 = \frac{1}{2}[m_A^2 + m_Z^2 \pm \root \of{(m_A^2 + m_Z^2)^2 - 4
m_A^2m_Z^2\cos^2 2\beta}] ,
\label{eq:mssmtl1}
\end{equation}
\begin{displaymath}
m_{H^\pm}^2 = \MW^2 + \MA^2 ,
\end{displaymath}
where \MW\ and \MZ\ are the W and Z masses, respectively, and \tanb\ is the ratio of the vacuum expectation values of the two 
Higgs fields. 
\par
The MSSM model may be constrained by the assumption that the sfermions (scalar fermions) masses, the gaugino masses and the 
trilinear Higgs-fermion couplings must unify at the Grand Unification scale (GUT).
In one of the possible constrained models the parameters chosen are:
\begin{itemize}
\item \MSUSY , a common mass for all sfermions at  the electroweak scale.
\item \Mdue , a common  $\mathrm SU(2)_L$ gaugino mass at  the electroweak scale.
\item $\mu$, the strength of the supersymmetric Higgs mixing.
\item \tanb , the ratio of the vacuum expectation values of the two Higgs fields .
\item A = $\rm{A_t}$= $\rm{A_b}$ a common trilinear Higgs-squarks coupling at the  electroweak scale. It is assumed to be the 
same for up-type squarks and for down-type squarks.
\item \MA , the  mass of the CP-odd Higgs boson.
\item \Mg , the gluino mass. 
\end {itemize}

\par Three of these parameters define the stop and sbottom mixing parameters $\rm{X_t}=\rm{A_t}-\mu~ \rm{cot}\beta $ and 
$\rm{X_b}=\rm{A_b}-\mu~ \rm{cot}\beta $.
\par
Whereas the particle spectrum depends on all the parameters mentioned above, the Higgs sector depends, at tree level, on only
two parameters which can be taken to be \tanb\ and \MA, as in Eq. (\ref{eq:mssmtl1}). The other parameters only enter through 
radiative corrections, but change
the mass prediction of Eq. (\ref{eq:mssmtl1}) (where it is limited to $\Mh < \MZ $), allowing the mass of h to reach higher 
values ($\approx 130$ GeV in some scenarios, see Sec. \ref{sec:Signal}).
\par 
Among all possible CP-conserving benchmark scenarios the so-called $\Mh-{max}$ scenario (Table \ref{tab:bensce}) has been 
considered \cite{Schael:2006as}.
This scenario corresponds to the maximum value of the stop mixing parameter $\rm{X_t}$=A - $\mu~ \rm{cot}\beta$ =  2\MSUSY.
Here the theoretical bound on the mass of the h is highest (hence the scenario's name) and experimental 
limits are less constraining. Also, the range of excluded \tanb\ values for given values of \Mt\ (the top mass) and \MSUSY\ 
is the most conservative one.
\begin{table}[ht]
\begin{center}
\begin{tabular}{|r||r|}
\hline
Parameter  & $\Mh-{max}$ \\
\hline
\MSUSY[\GeV] & 1000  \\
$\mu$ [\GeV]& -200  \\
\Mdue [\GeV] & 200  \\
$\rm{X_t}$=A - $\mu~ \rm{cot}\beta$ & 2\MSUSY \\
\Mg[\GeV] & 0.8\MSUSY  \\
\hline
\MA[\GeV] & 0.1-1000  \\
\tanb\ & 0.4-50  \\
\hline
\end{tabular}
\caption{CP-conserving benchmark $\Mh-{max}$ scenario .}
\label{tab:bensce}
\end{center}
\end{table}
\par
We shall focus on the  $\Mh-{max}$ scenario as the most promising for the search of the \h\ boson, referring to it hereafter 
as MSSM. 
\par
At tree level the MSSM Higgs boson couplings to fermions and massive gauge bosons are obtained from the SM Higgs boson couplings  
via correction factors \cite{ Richter98}. They depend on the parameters $\beta$ (already introduced) and $\alpha$, the mixing 
angle which diagonalizes the CP-even Higgs boson mass matrix. The two parameters are related by the following expression:
\begin{equation}
\cos 2\alpha = - \cos 2\beta \ \frac{m_A^2 - m_Z^2}{m_H^2 - m_h^2}
\end{equation}
\par
At high \tanb\ the MSSM correction factors to the SM Higgs bosons couplings to fermions and massive gauge bosons are larger 
for down-type quarks (b) and leptons ($\tau$ and $\mu$) than for up type-quarks. Thus, the associated $\bb$h production is 
enhanced and becomes the dominant process in the production of h bosons in the high \tanb\ region. 
\par
\begin{figure}[h]
\begin{center} 
\epsfig{file=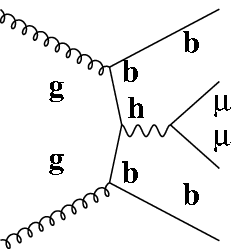,width=50mm}
\hspace{1cm}
\epsfig{file=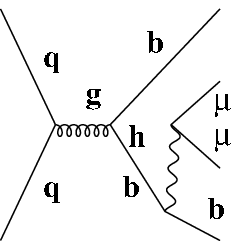,width=50mm}
\end{center}
\hspace{0.5cm}
\caption{  Typical diagrams contributing at ``tree level'' to the process
\gg$\to$ \BBh and \qq $\to$  \BBh.}
\label{fig:ggqqh}
\end{figure}

The Feynman diagrams contributing to the process \gg$\to$ \BBh and \qq $\to$  \BBh\ are shown in Fig.~\ref{fig:ggqqh}. 

\hspace{1cm}
\par
We shall consider mainly the decays to \mm. Indeed, although the Higgs boson couplings are proportional to the fermion mass, thus 
resulting in a branching ratio to \TT\ higher than to \mm\  by a factor $(\frac{m_{\tau}} {m_\mu})^2$, the experimental conditions 
favor the \mm\ channel\footnote{
The production advantage of the \TT\ channel is counterbalanced by the difficulty of identifying the hadronic decay of a $\tau$-jet 
in hadronic events, by a smaller acceptance of the detector and by a worse mass resolution due to the presence of neutrinos in the 
final state. Instead, with a final state like $\h\ra\mm $ ATLAS would exploit the excellent combined performance of the muon 
spectrometer and inner detector.}. 
\par
In the region of  high \tanb~ and $\Mh \approx 100$ \GeV\ the CP-odd neutral Higgs boson A has a mass only slightly 
higher than the CP-even h and a competitive branching ratio for the \mm\ decay channel \cite{ Richter98}. Also, cross-sections and 
widths, which  are functions of  the parameters \tanb~ and \MA, are close in some regions of the parameter space 
(Sec. \ref{sec:Signal} ). 
Thus, in these regions the h and A bosons are indistinguishable from an experimental point of view, and it is more correct 
to think in terms of a h/A search. In the following we refer to the boson sought as the h/A boson (however its mass is noted \Mh\ or \MA, 
accordingly) . The degeneracy between h and A is less pronounced near the higher mass limit of h .

%% file: Exp.tex
\section{Experimental search for  Minimal Supersymmetric Standard Model Higgs}\label{sec:Exp}
\subsection{LEP and Tevatron results }\label{sec:LEP}
High precision tests of the Standard Model have been performed at LEP setting a combined limit of \MH $ >$ 114.4 \GeV\   
for the mass of the SM Higgs boson \cite{Barate:2003sz}. 
\par
\begin{figure}
\begin{center}
\begin{tabular}{cc}
\includegraphics*[width=0.45\textwidth]{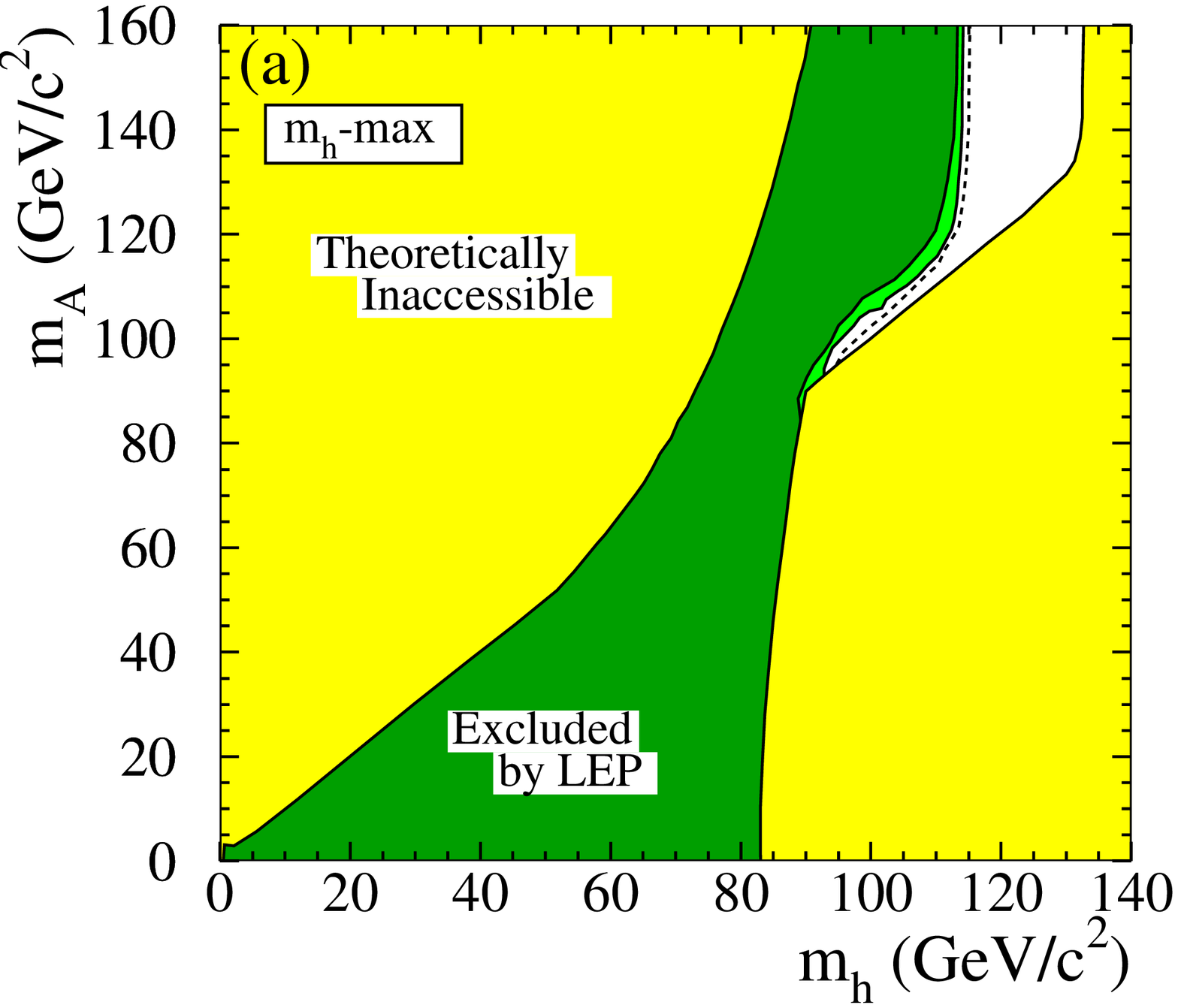} &
\hspace{-0mm}
\includegraphics*[width=0.45\textwidth]{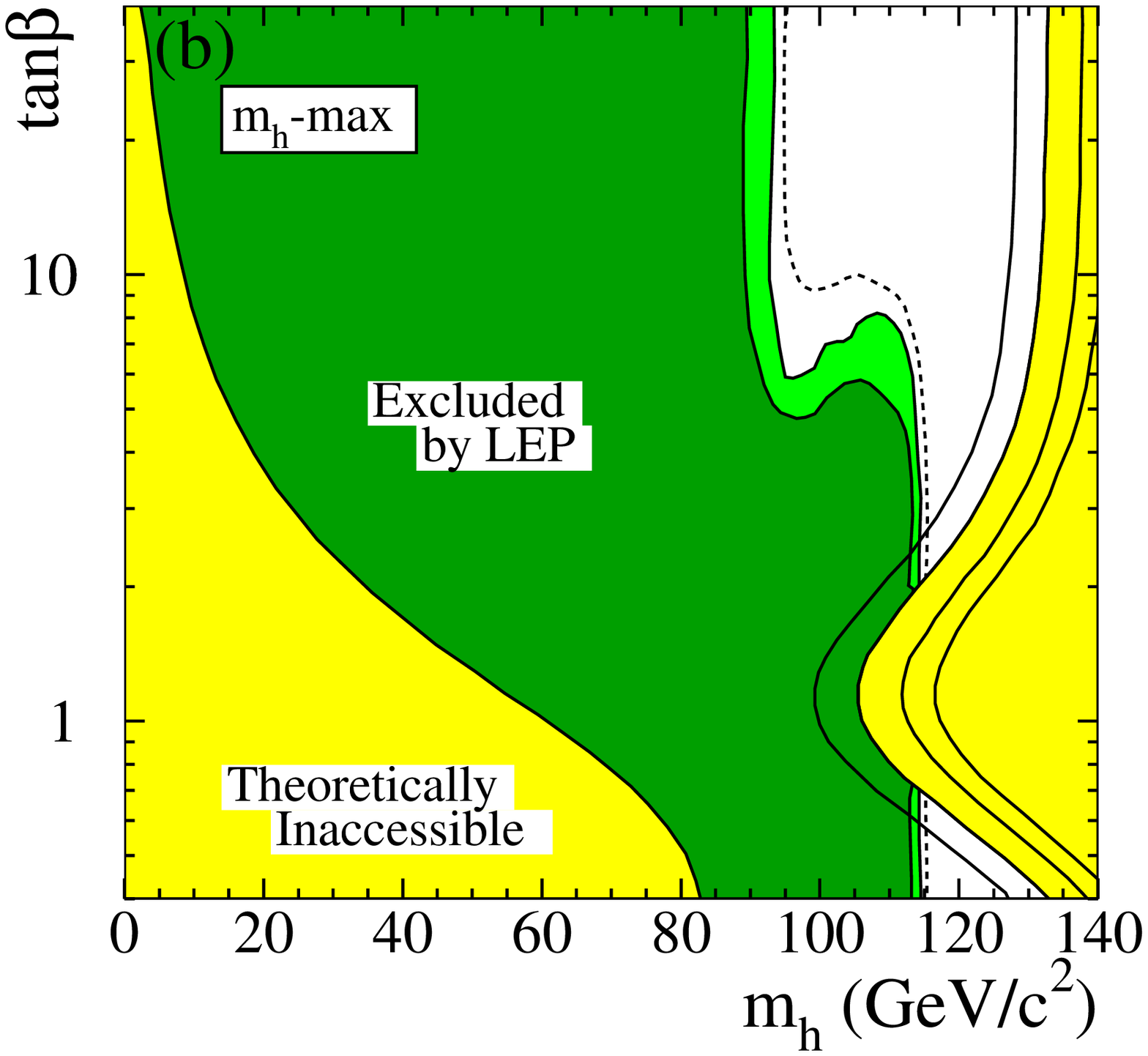} \\
\end{tabular}
    \caption[]{\label{fig:minmax}
             The combined LEP results for the search for the MSSM neutral Higgs bosons (from Ref.~\cite{Schael:2006as}).
             The figure shows the theoretically inaccessible regions (light-grey/yellow) and the regions experimentally excluded by
             LEP searches, at 95\% C.L.~(medium-grey/light-green) and 99.7\% C.L.~(dark-grey/dark-green), for the 
             {\it \Mh-max} scenario
             with the top mass \Mt = 174.3 \GeV, in two projections of the MSSM parameters (\Mh,~\MA), (\Mh,~\tanb).
             The dashed lines indicate the boundaries of the regions which are expected to
             be excluded, at 95\% C.L., on the basis of Monte Carlo simulations with no signal. In the (\Mh, \tanb )
             projection, the upper boundary of the parameter space is indicated for four values of the top mass;
             from left to right: \Mt = 169.3, 174.3, 179.3 and 183.0 \GeV.
            }
\end{center}
\end{figure}

Again at LEP, the validity of the Minimal Supersymmetric Standard Model has been investigated within the constrained 
framework of Sec. \ref{sec:MSSM}. For the mass of the charged MSSM Higgs bosons a combined limit \MHPM $>$ 78.6 \GeV\  
was obtained  \cite{unknown:2001xy}. Searching for neutral CP-even and  CP-odd MSSM Higgs bosons, no indication of signal 
was found up to a center-of-mass energy of  209 \GeV\ \cite{Schael:2006as}.  The corresponding lower 
limits on the masses were set as a function of \tanb\ for several scenarios. In the {\it \Mh-max} scenario (Fig.~\ref{fig:minmax}) 
with a top mass \Mt =174.3 \GeV\  the limits for $\tanb > 10$ at 95\% CL are approximately :
\begin{displaymath}
\Mh , \MA  \geq  93~ \GeV
\end{displaymath}
  
\par
A complementary search, providing sensitivity in the region  $\tanb > 50$  has been performed at the Tevatron Collider at 
\RS = 1.96 \TeV. In the MSSM scenario, a significant portion of the parameter space  has been  excluded by the D0 Collaboration,
down to \tanb\ = 50 as a function of \MA, by studying the associated production with two b quarks of h/A/H bosons and their 
decay into \bb\ \cite{collaboration-2005-95}. Comparable results have been obtained by the CDF Collaboration exploring the 
h/A/H decays to \tautau, but extending the excluded region to higher values of \MA\ \cite{collaboration-2006-96}.

\subsection{LHC discovery perspectives}\label{sec:LHC}
The LEP and Tevatron data don't exclude the parameter space defined by \tanb\ larger than 10 and smaller than 50.
Therefore, a natural continuation of the  LEP  and Tevatron physics is the investigation of the possible existence of MSSM  
Higgs bosons in this region of \tanb.  
The ATLAS \cite{ATLAS} and CMS \cite{CMS} experiments starting in the near future at the  Large Hadron Collider (LHC), at CERN, 
constitute an excellent laboratory for such search. 
\par
The prospect for the detection of MSSM Higgs bosons at LHC was evaluated for benchmark sets preventing Higgs boson decays to 
SUSY particles \cite{Richter98,ATLAS} and  focusing on the discovery potential of decay modes common to MSSM and SM Higgs bosons 
\cite{ATLAS}. It was concluded that the complete region of parameter space \MA\ = 50 -- 500 \GeV\ and \tanb\  = 1 -- 50 is 
open to Higgs boson discovery by the ATLAS experiment, already with an integrated luminosity of \IL\ = 30 \fbinv, and that  
over a large part of this region more than one  Higgs boson and more than one decay mode could be observed --
the detection of a signal in more than one decay channel would constitute strong evidence for the MSSM model.
It was also found that the region in the (\MA, \tanb ) plane which corresponds to \Mh\ $\approx$ 100 \GeV\ 
and $\tanb\ > 10$ is only accessible by a neutral h/A boson decaying to \mm\ or \tautau\ \cite{Richter98,ATLAS},  and by a 
charged \Hpm\ boson decaying to $\tau\nu $ \cite{Assamagan:2002ne}. 
\par
More recently the h boson discovery potential in the MSSM scenario has been investigated \cite{Schumacher:2004da} at two 
luminosities, \IL = 30 \fbinv\ and \IL = 300 \fbinv\ .
At low luminosity the  $\h \ra\TT$ decay mode represents the main contribution to  the discovery potential and covers 
most of the parameter space not yet explored. However the contribution of  \bb$\h \ra\mm$ appears to be crucial in the 
region of moderate \tanb\ and mass close to \MZ. The channel  $\bb \h \ra \mm$  which requires an excellent performance 
in $\mu$ detection and  b-tagging is well suited to the ATLAS experiment thanks to a design giving high performance from 
the muon spectrometer and the inner detector. 

At high luminosity channels such as: $\h \ra\gamma\gamma$,  $\h \ra\Zo\Zo \ra 4 \ell$ and $\h \ra\bb$  in associated 
production with  \tt\  give a significant contribution. 
The channel $\h \ra\gamma\gamma$, which requires an excellent $M_{\gamma\gamma}$  mass  resolution and jet/$\gamma$ separation, 
corresponds to MSSM rates suppressed with respect to the SM case but for a limited region of the parameter space where they 
could even be slightly enhanced. 
As for the channel $\h \ra\bb$, only the \tt h production followed by the $\h \ra\bb$ decay can be observed clearly above 
the background, thus the extraction of the signal requires the identification of four b-jets and an excellent b-tagging 
performance. In the MSSM case the rates could be enhanced by 10-20\% over the SM rates.

\subsection{Background processes}\label{sec:Background}
The prospects for the detection of MSSM  Higgs bosons at LHC depend heavily on the suppression of the background sources:
\begin{itemize}   
\item $\Zo/\gamma^*$ production with two b-jets and a subsequent decay into a \mm\ pair.
The cross section for this process is a few orders of magnitude larger than for the signal. As an example, we quote for the 
\Zo\ boson $\sigma_{\bb\Zo} \cdot \rm {Br_{\Zo \ra\mm}}\approx$ 22.8 \pb\ \footnote{Evaluated from AcerMC(2.3) \cite{Kersevan:2004yg} 
and PYTHIA 6.226 \cite{Sjostrand:2004yu} (with  \MZ~ $>$ 60 \GeV ).} and for each of the h and A bosons $\sigma_{\h / \Aa\bb} \cdot 
\rm{Br}_{\h / \Aa \ra \mm} 
\approx  \ 0.24$ \pb\ (at \tanb\ =\ 45 and \Mh\ =\ 110~\GeV )\footnote{Evaluated from PYTHIA 6.226 \cite{Sjostrand:2004yu}.}.
\begin{figure}[h]
\begin{center}
\epsfig{file=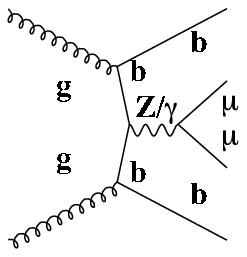,width=50mm}
\hspace{1cm}
\epsfig{file=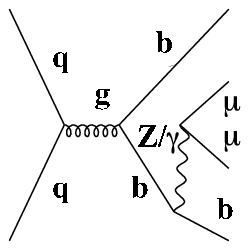,width=50mm}
\end{center}
\caption{ Typical diagrams contributing  at ``tree level'' to the process
\gg $\to \Zo/\gamma^* \bb \to \mm \bb$ and $\qq~\to~\Zo/\gamma^*\bb~\to~\mm\bb$.}
\label{fig:ggqqz}
\end{figure}
The corresponding diagrams are shown in  Fig.~\ref{fig:ggqqz} and Fig.~\ref{fig:ggqqh}, respectively. It is clear that they 
differ only in the kind of boson produced, h (or A) in the signal and $\Zo/\gamma^*$ in the background. The distinction between 
the signal and the background when \Mh\ is approaching  \MZ\ becomes then an extremely hard task due to the similar topology of 
the decays \cite{Gentile07}, although angular distributions differ somewhat due to the fact the boson is a scalar in one case 
and a vector in the other.
\item
\Zo\  production with two jets not originating from b-quarks. The cross section  is $\approx 24$ times that for Z with two 
b-jets\footnote{Evaluated (for this purpose) from bbZ and jjZ cross sections by SHERPA 1.0.9 \cite{Gleisberg:2003xi} (with $\MZ >60$ 
GeV).} and can contribute to the background in case of jet misidentification. An estimate of a possible impact on the significance
of this analysis is reported in Sec. \ref{sec:RefPoint}. 
\item 
\Zo\Zo\ associated production, when one \Zo\ decays into \bb\ and the second one decays into \mm . This process has a cross section 
of the same order of magnitude as the signal, $\sigma_{\Zo\Zo } \cdot \rm {Br_{\Zo \ra \mm}}\cdot \rm {Br_{\Zo \ra \bb}}\approx \ 
0.13~\pb $~\footnote{Evaluated from PYTHIA 6.226 \cite{Sjostrand:2004yu}.}, but can be easily suppressed using the kinematic 
characteristics of the events, see following sections. 
\item 
\tt\ associated production followed by a top-quark decay into a b-quark and a W boson and a subsequent W decay in $\mu \nu$. 
The cross section of this process is $\sigma_{\tt } \cdot  {\rm Br_ {t \ra b\W}}\cdot {\rm Br_{W \ra \mu\nu}}\cdot 
{\rm Br_{t \ra b\W}}\cdot {\rm Br_{W \ra \mu\nu}} \approx  \ 5.71~ \pb $~\footnote{Evaluated from PYTHIA 6.226 
\cite{Sjostrand:2004yu}.}. The presence of two neutrinos 
implies missing transverse energy in the event (see following sections) thus allowing this background to be strongly reduced. 
One would want to discriminate the signal from the background  on the basis of the different b-jets characteristics (the two
background b-jets are usually more energetic than those accompanying the signal and the probability of their identification is 
higher), but the requirement that the two b-jets be identified will suppress the signal more than the background.
\end{itemize}

%% file: Simul.tex
\section{Monte Carlo simulation}\label{sec:MC}
We improve on the analyses reported in Sec. \ref{sec:LHC} with a full Monte Carlo simulation of the experiment 
(data generation, reconstruction  and analysis). The exploration of the unexcluded MSSM parameter space, with a view 
to either discovering a supersymmetric Higgs boson or excluding the model considered, constitutes the motivation of the 
analysis described in this paper.
\par
The considerations of Sec. \ref{sec:MSSM} and Sec. \ref{sec:Exp} suggested a search for $\h /\Aa \ra \mm$ decays 
accompanied by two b-jets. The extraction of the h/A boson signal from  the competing  enormous background of Z decays in 
the mass region close to \MZ\ constitutes a challenging search, where all performances of the experimental setup  have to 
be exploited.
\par
To this purpose we have generated signal and background Monte Carlo events using the PYTHIA program (v.6.226)
\cite{Sjostrand:2004yu} as specified below, for a center-of-mass energy  \RS = 14 \TeV. The efficiency of the selection 
criteria, the  detector acceptance and the purity of the data sample are estimated from these events and the ATLAS detector 
response \cite{ATLAS}. The latter is simulated using the GEANT program \cite{Geant03,Geant06} which takes into account the 
effects of energy loss, multiple scattering and showering in the detector through an interface called ATHENA (v.10.0.1).
\par
The number of events used in this analysis corresponds to an integrated luminosity $ \IL =  300~ \fbinv$ with the 
exception of the channel \BBZ. In this latter case, for practical reasons, a number of events corresponding to half the 
mentioned luminosity has been generated. We note that $ \IL =  300~ \fbinv$ corresponds to the 
integrated luminosity expected after three years of data taking.

\subsection{Signal}\label{sec:Signal}
The MSSM neutral Higgs bosons h, A and H were generated in associated production with two b-quarks using the PYTHIA program 
(v.6.226) through the ATHENA interface (v.9.0.4). The parameters of the model were given the values \cite{Mrenna05} shown 
in Table \ref{tab:Pythia_params } (for \tanb\ and \MA\ the range of values scanned is shown). In this parameter region, as 
mentioned in Sec.~\ref{sec:MSSM}, the cross section, mass and width of h and A bosons are close while the H cross section 
is one order of magnitude lower than the  h/A cross section. 

As is done in the Pythia code, we have taken \MA\ as input parameter.  The values of \Mh\ and \MH\ are then derived 
from PYTHIA as a function of \MA.

\begin{table}[h]
\begin{center}
\begin{tabular}{|c|r|r|}
\hline
Parameter& &value \\
& & \\

\hline
common gaugino mass& \Mdue$\equiv m_{1/2}$~[\GeV]& 200 \\
\hline
gluino mass&\Mg~[\GeV] &800  \\
\hline
strength of supersymmetric Higgs &$\mu$ [\GeV]& - 200 \\
\hline
ratio of Higgs fields&\tanb & 15-50   \\
\hline
common scalar mass &$m_0$~[\GeV] & 1000 \\
\hline
squark left 3 gen& $M_{\tilde{\rm qL}}$~[\GeV]   &1000    \\
\hline
sbottom mass& $M_{\tilde{\rm bR}}$~[\GeV]   &1000    \\
\hline
stop mass  &$M_{\tilde{\rm tR}}$~[\GeV] &1000  \\
\hline
stop-trilinear coupling  &  A &2440  \\
\hline
 mass of CP-odd boson &\MA~ [\GeV]& 95-135  \\

\hline
\end{tabular}
\end{center}
\caption{ Parameters for A, h and H generation in PYTHIA (v.6.226)~\cite{Sjostrand:2004yu}.  }
\label{tab:Pythia_params }
\end{table}

\par
The signal  \BBh\ has been simulated in 104 points of the parameter space (\MA, \tanb ), corresponding to eight 
steps in \tanb, chosen equally spaced  between 15 and 50  and thirteen steps of 2.5 \GeV~  in  \Mh~   between 95 
\GeV\ and 125 \GeV\ (the largest value allowed by PYTHIA (v.6.226)). These are also the points where the decay \BBA\ 
has  been  simulated. 
\par 
The values of the mass, cross section and width of the  A/h and H neutral bosons are reported in Ref. \cite{Gentile07} 
for all the points analyzed of the (\MA, \tanb ) plane. For the following discussion a reference point has been chosen 
at \tanb\ = 45,~\mbox{\MA=110.31 ~\GeV } (\Mh = 110 \GeV, \MH = 127.46 \GeV). The corresponding width is \GA = 4.28 \GeV\
(\Gh = 4.20 \GeV, \GH = 0.05 \GeV), while the production cross section times the branching ratio for the decay to \MM\ is 
$\sigma_{\BBA} = 0.243$ pb ($\sigma_{\BBh} = 0.245$ pb, $\sigma_{\BBA} = 0.0016$ pb).    

\subsection{Background}\label{sec:Back}
For convenience we list again below the sources of background considered in Sec. 3.3 which were fully simulated. The Z
production associated with two jets not originating from b as yet has not been implemented in the AcerMC2.3 code and has 
been ignored at this level (see however end of Sec. \ref{sec:RefPoint}).
\begin{itemize}
\item \BBZ.
The events are generated by AcerMC(2.3) \cite{Kersevan:2004yg}  with the hadronization process of  PYTHIA (v.6.226)
\footnote{Using PHOTOS package for {\it inner} \brem~  generation.}. At the  generator level a cut-off is applied to the \mm\  
invariant  mass, $\Minv  > 60~\GeV $ \footnote{ A low energy cut is fixed on {\it inner} \brem\ photons at $\PT > 5 ~\GeV$.}. 
\item \BBTT.
The events are fully generated with PYTHIA (v.6.226).
\item \ZZ.
The events are fully generated with PYTHIA (v.6.226). 
\end{itemize}

\begin{table}[h]
\begin{center}
\begin{tabular}{|c|r|r|r|r|}
\hline
process& $\sigma_{\bb\mm }$&$\rm N^{\rm exp_{300}}$ &$\rm N^{\rm MC}$ &${\rm w}$ \\
(background) &~~[pb]~~~~&& & \\
\hline
\BBZ &  22.789     & 6836700 &3314000 & 2.06 \\
\hline
\BBTT&   5.71      & 1713420 &1806437 & 0.95 \\
\hline
\ZZ  &   0.1273    & 33819 & 97244    & 0.35 \\

\hline
\end{tabular}
\end{center}
\caption{Background cross section times branching ratios, $\sigma_{\bb\mm}$, number of expected events for 
         $\IL = 300~ \fbinv $, $\rm N^{\rm exp_{300}}$, number of Monte Carlo generated events, $\rm N^{\rm MC}$, 
         and their weight in the analysis, w, for the three processes considered. 
         }
\label{tab:background}
\end{table}
\par 
The cross section times branching ratios, together with the number of expected events for $ \IL = 300 ~\fbinv$, the 
number of  Monte Carlo generated events and  their weight, is reported  in Table \ref{tab:background}  for the three background 
processes. 
\par
Among these the main contribution comes from \BBZ, and is affected by a cross section uncertainty arising from 
uncertainties on QCD, QED couplings  ($\approx 10\%$ \cite{Kersevan:2004yg}) and from  higher order corrections 
($\approx 25\%$ \cite{Campbell:2003hd}). As we intend to account for this large uncertainty with a data-driven method  
(Sec.~\ref{sec:Prelim3}), a sample of $\bb\Zo \ra \ee$ events, $\approx$ 600000 events, corresponding to an integrated luminosity 
$ \IL =  30~ \fbinv$ with a cross section $\sigma \approx 22.8~ \pb$, has also been simulated. Owing  to the large number of 
events simulated, there is only a minor statistical uncertainty associated with the evaluation of the tails of $\Zo\ra\mm $ 
background. 

\section{The ATLAS detector}\label{sec:det}

Here we shall summarize the main features of the ATLAS detector ~\cite{ATLAS1} which are relevant for the present analysis.
For a more detailed information we refer to  \cite{Froidevaux:2006rg} :

\begin{itemize}
\item {\bf Magnet System}:
This consists of  a solenoid providing a 2 T$\cdot$m  bending power in the inner detector, and a barrel air toroid completed by two 
end-cap toroids, with a typical bending power of 6~T$\cdot$m and  3~T$\cdot$m respectively.
\item {\bf Inner Tracking System}: 
It has been designed to measure as precisely as possible and with high efficiency  the charged particles emerging from primary 
interactions in a pseudorapidity range  $\left| \eta \right| < 2.5  $.
It is installed inside in the solenoid magnetic field and consists of  Si pixels and silicon strip detectors near the interaction 
point, and strawtubes. 
The expected transverse momentum resolution of the inner tracker for a 100 \GeV~  charged particle at $\left| \eta \right| =0  $  
is $\approx  3.8\%$.
\item {\bf Electromagnetic Calorimeter System}: 
The excellent energy resolution and particle identification for electron, photons and jets demanded from physics is realized
from a  liquid argon-lead  sampling calorimeter with accordion shape in the barrel and end-cap regions.
The energy resolution  expected  is 
 $\frac{\sigma}{\rm{E}}=  \frac{10-12\%}{\sqrt{\rm{ E}}}\oplus 0.35$.
\item {\bf Hadronic Calorimeter}: 
It is a copper-liquid argon calorimeter  in  the end-cap region  and a Fe-scintillator calorimeter  scintillators in barrel region. 
The liquid argon tungsten forward calorimeters extend the coverage to $\left|\eta \right|$ =4.9. The  expected  hadronic jet energy 
resolution is  $\frac{\sigma}{\rm{E}}=\frac{4.7}{E}\oplus \frac{55\%}{\sqrt{\rm{ E}}}\oplus 0.013$. 
\item {\bf Muon spectrometer}. 
The  reconstruction of muons at highest luminosity is one of the most important point in ATLAS design.
The ATLAS toroidal magnet field provides a muon momentum resolution that is independent of pseudorapidity.
The  spectrometer is  constitued by:
a) The precision tracking chambers made of  Monitored Drift Tubes (MDT) covering the rapidity range $\left| \eta \right| < 2.7  $,
b) The Cathode Strip Chambers (CSC) and transverse coordinate strips cover the most forward rapidity region (  $\left|  \eta \right| 
= 2.0- 2.7 $) in the  inner most stations of the muon system. 
c)  Resistive Plate Chambers (RPC) in the barrel region (  $\left| \eta\right|  < 1 $) and  Thin Gap Chambers (TGC) in the end-cap 
region  provide  muon triggers and measure  the second coordinate of the muon tracks. 
The expected momentum resolution ranges from about 1.4\% for 10 GeV muons to 2.6\% for 100 GeV muons  at$\left| \eta \right| =0  $
\end{itemize}
\section{Preliminaries to Monte Carlo data analysis}\label{sec:Prelim}

Two points are crucial for our analysis:
\begin{itemize}
\item The muon reconstruction efficiency in the  analysis acceptance and the \mm\ invariant mass resolution.
\item The b-jet identification.
\end{itemize}

\subsection {Efficiency and resolution studies with ${\boldsymbol \BBZ }$ events} \label{sec:Zmm}
The  $\mu$ reconstruction performance of the apparatus was studied with  a sample of \BBZ\ events \cite{Pomarico05}.
For this purpose only events with two reconstructed muons of opposite charge found within the $\vert \eta \vert~ \leq ~2.5 $ 
acceptance were considered. The contribution of tracks mimicking muons at reconstruction level was found to be negligible in a
\Zo\ra\mm\ sample \cite{Adams06}. The pileup effect seems to be also negligible based on the information presently available
\cite{Rosati05}.
\par
\begin{figure}
\begin{center}
\begin{tabular}{cc}
\hspace{-7mm}
\includegraphics*[width=0.45\textwidth]{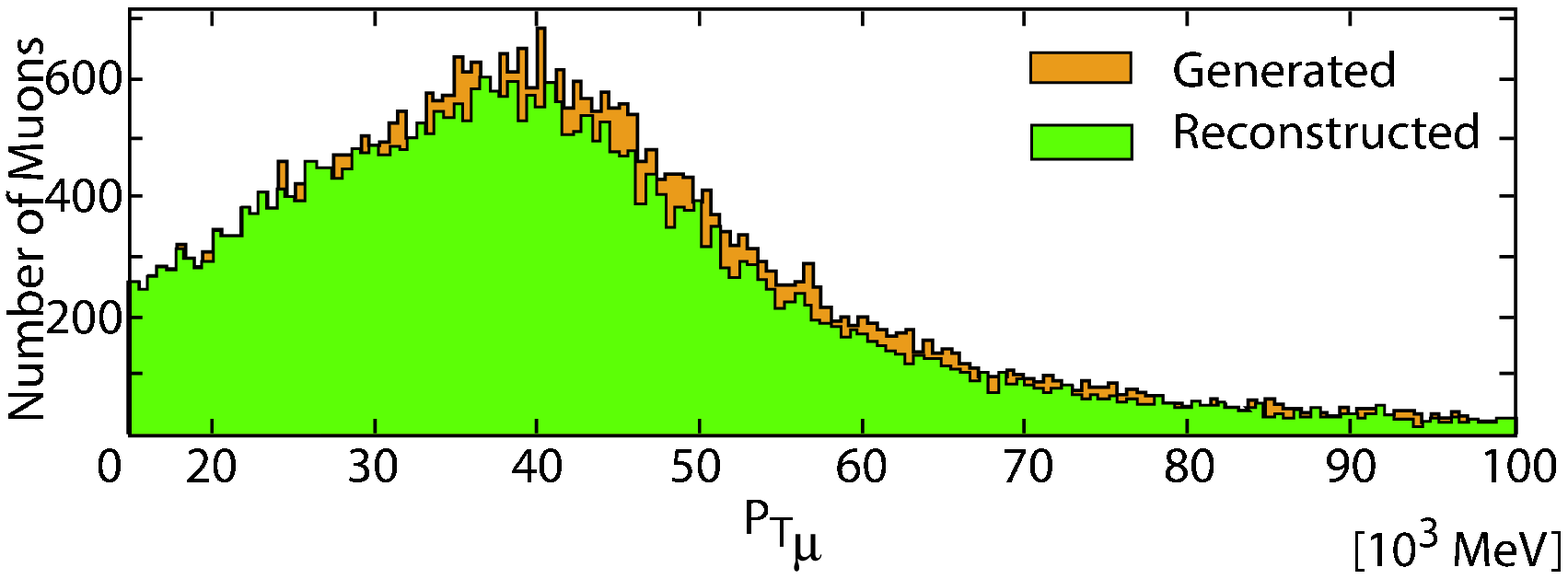} &
\hspace{-0mm}
\includegraphics*[width=0.45\textwidth]{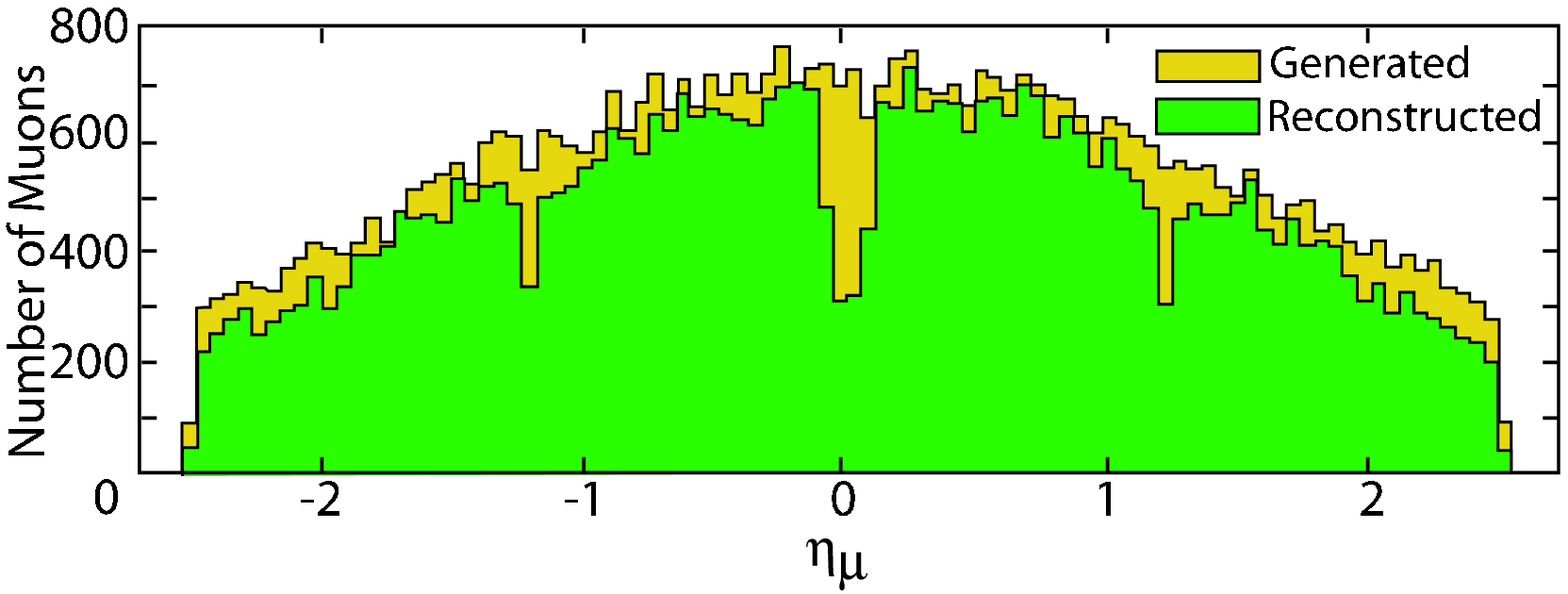} \\
\hspace{-7mm}
\includegraphics*[width=0.45\textwidth]{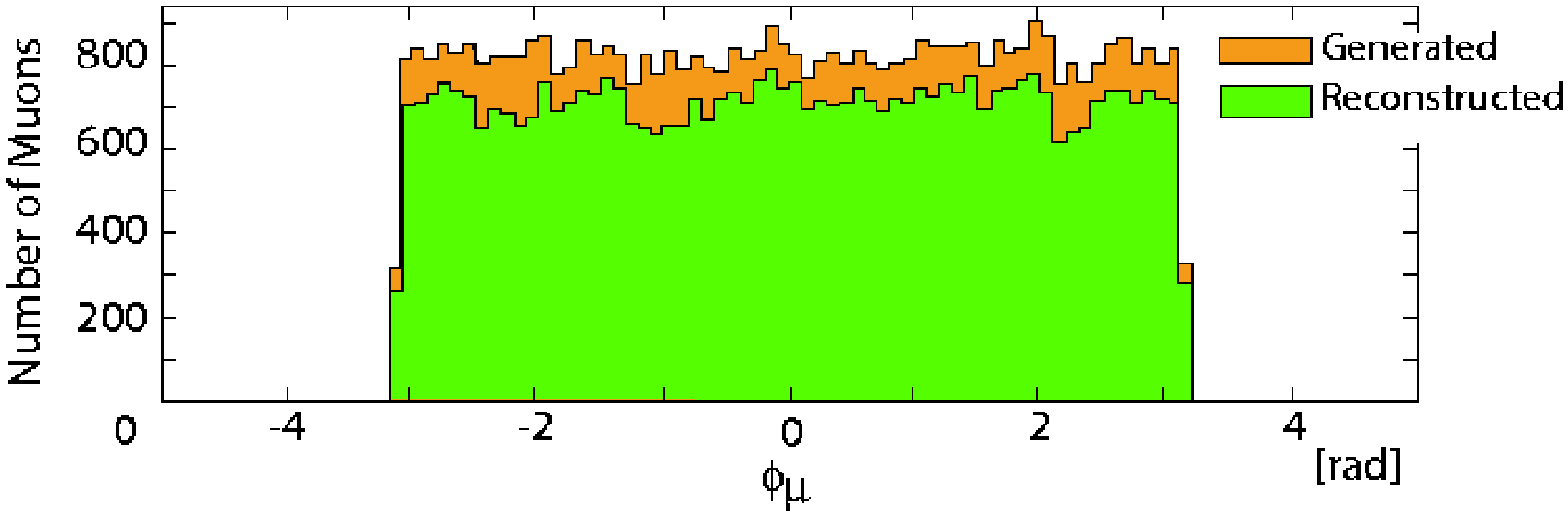} &
\hspace{-0mm}
 \\
\end{tabular}
    \caption[]{\label{fig:Muons_distr_pt}
            Distributions of the transverse momentum \PTmu, pseudorapidity $\eta_\mu$ and polar angle $\phi_\mu$ for muons
            from \bb\Zo ($\ra$ \mm ) Monte Carlo events, generated (brown) and both reconstructed (green) within the 
            $\vert \eta \vert~\leq ~2.5 $ acceptance.     
            }
\end{center}
\end{figure}

The  distributions of the transverse momentum, \PTmu,  pseudorapidity, $\eta_{\mathrm{\mu}}$, and polar angle, 
$\phi_{\mathrm{\mu}}$, of reconstructed muons (see Fig. \ref{fig:Muons_distr_pt}, green histograms) reproduce with good efficiency 
the generated data (see Fig. \ref{fig:Muons_distr_pt}, brown histograms).

\par
\begin{figure}
\begin{center}
\hspace{-7mm}
\includegraphics*[width=0.60\textwidth]{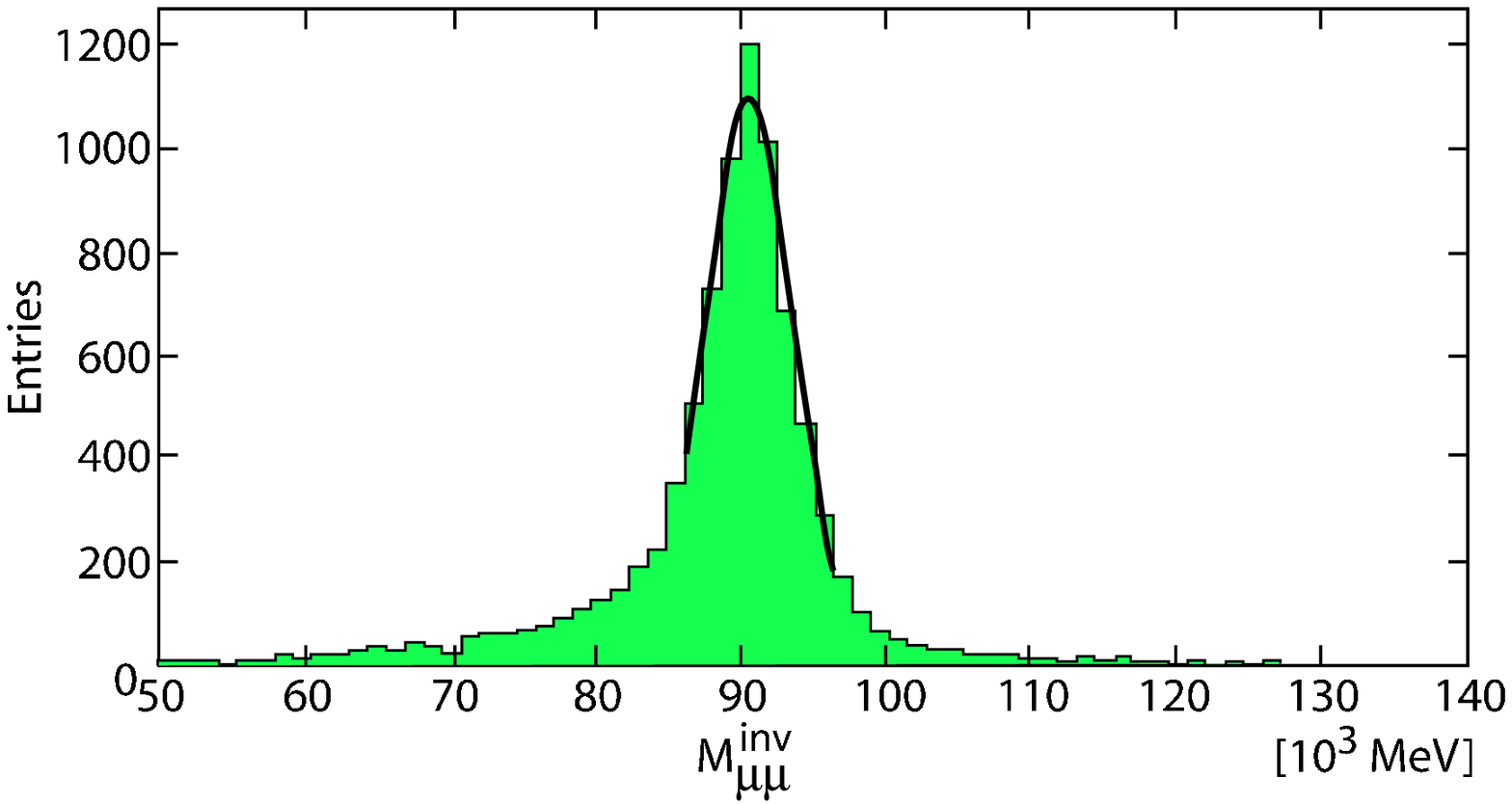} \\
\includegraphics*[width=0.60\textwidth]{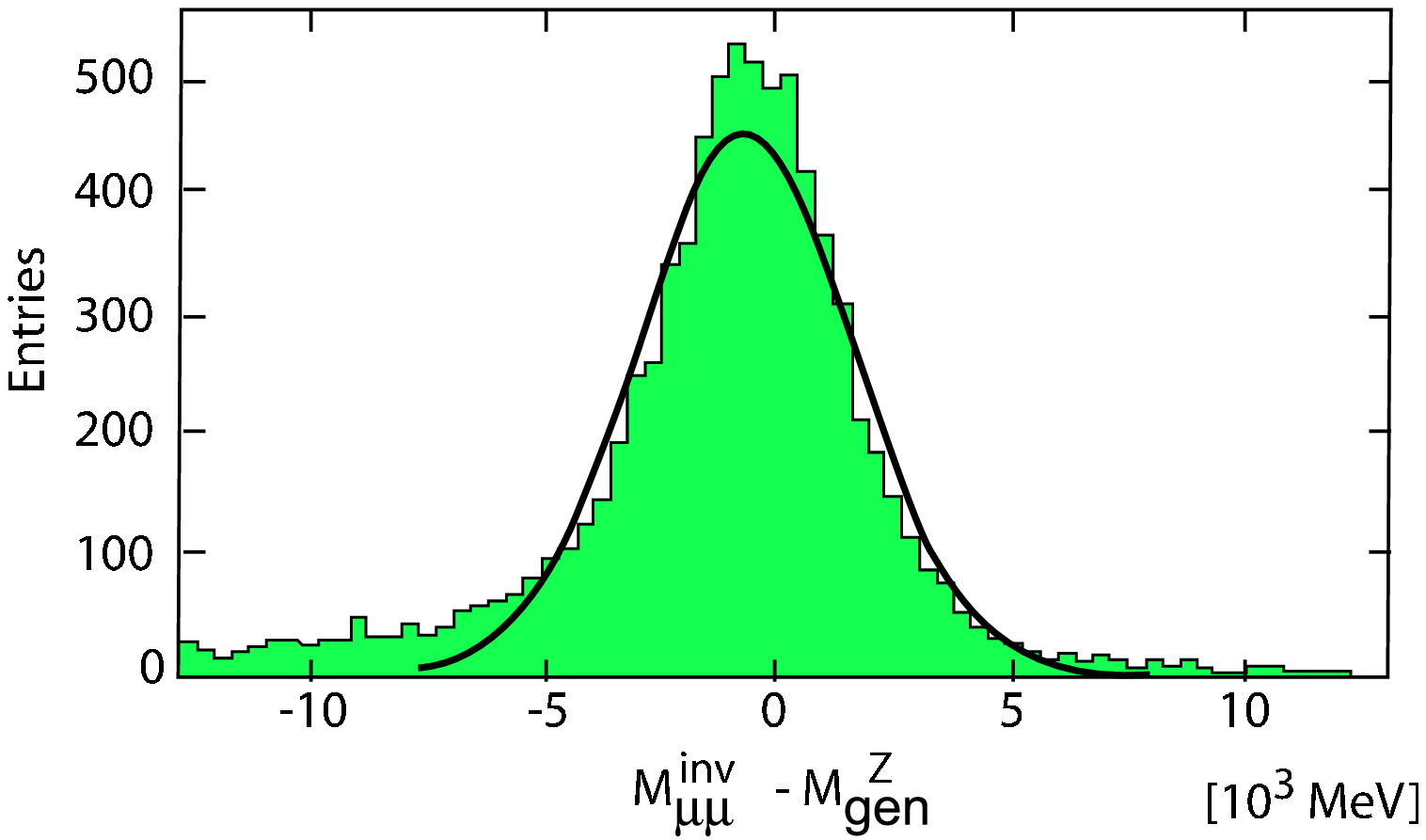}
 \caption[]{\label{fig:Minv_z}   
         (top) Distribution of the invariant  mass \Minv\ reconstructed  in \bb\Zo ($\ra$ \mm ) events. 
         (bottom) Distribution of the difference between reconstructed and generated mass 
          $ \Minv -  M^{\Zo}_{\rm gen }$ for the same sample of events.
         }
\end{center}
\end{figure}

  
The distribution of the reconstructed dimuon invariant mass is shown in Fig.~\ref{fig:Minv_z} (top) together with the result
of a Gaussian fit. The solid line corresponds to the following values for the fit parameters:

\begin{equation}
< \Minv > = 90.47 \pm 0.05 ~\GeV ~~~~~
\sigma (\Minv ) = 3.02 \pm 0.06~\GeV \\
\label{eq:Mz}
\end{equation}

The fit mean is smaller  than the nominal value of  \MZ\ \cite{Yao:2006px}. The \Zo\ natural width contributes to $\sigma (\Minv )$ 
for approximately 1.9 \GeV\, thus implying a measurement accuracy $\sigma _{\rm res} = 2.3 ~\GeV $. This is just the value obtained 
unfolding the reconstructed distribution with the distribution of the \Zo\ generated mass, $M_{\rm gen }^{\Zo}$. When the difference 
between reconstructed and generated mass $\Minv\ -  M_{\rm gen }^{\Zo}$ is plotted, Fig.~\ref{fig:Minv_z} (bottom), the Gaussian fit 
of the distribution gives the following results for the fit parameters:
\par 
\begin{equation}
< \Minv - M_{\rm gen}^{\Zo} > = -0.82 ~ \pm ~ 0.03~\GeV ~~~~
\sigma _{\rm res}= 2.35~ \pm~ 0.03 ~\GeV \\
\label{eq:Mz_fit }
\end{equation}

\par 
From this study we conclude that in the \MZ\  region the reconstructed \mm\ invariant mass distribution shows a mean value 
shifted by 820 \MeV\ with respect to the nominal value, towards the low mass region, and a resolution of $\sim$ 2.6 \%. 
These results will be used  later in  Sec.~\ref {sec:RefPoint}. 

\subsection{ b-tagging studies} { \label{sec:btag} 
The two b-tagging algorithms called 3D and SV2, which are in use within the ATLAS Collaboration \cite{Kostyukhin03}, were 
studied with a subsample of \BBh\ events.

Starting from the impact parameters (transverse and longitudinal) and their significances,  both algorithms assign a weight 
to each track of the jet, that is to each track within a cone of opening angle $\DR = \sqrt {\Delta \eta ^2 + \Delta \phi ^2}$,  
with at least one track reconstructed in the tracker. The track weight is the ratio between the likelihood functions for a
b-jet track and for a track of a light jet (from a light quark, u, d, s, c  or a gluon). In turn the jet weight 
$\mathrm{w_{jet}}$ 
is defined as the sum of the logarithms of the tracks weights. This allows b-jets $(\rm w_{jet} > \rm w_{cut} )$ to be 
discriminated against light jets ($\rm w_{jet} < \rm w_{cut}$) -- 
the value $\rm w_{cut}$ is chosen, depending on the physics under study, such as to have a high 
efficiency  $\epsilon_{\rm b}$ for b-jet identification and a high rejection factor $\rm {R_j}$ for light jets.

The SV2 algorithm improves on the 3D algorithm by using  additional information on other variables not too strongly correlated 
with the impact parameters (as \eg\ fraction of the jet energy and invariant mass of all particles at the secondary vertices, 
number of two-track secondary vertices).

\par In Table \ref{tab:b_tag }, the results of a study of a \BBh\ sample are reported. Similar results have been  obtained 
from a  
\BBZ\ sample. For both samples, a selection cut, $\PTjet > 15~ \GeV$, is applied, since below this value the efficiency of 
b-identification drops. The study is limited to the inner detector acceptance $\vert \eta \vert \leq 2.5$ and to two values 
of the jet opening angle, $\DR = 0.7$ and $\DR = 0.4$

\begin{table}[h]
\begin{center}
\begin{tabular}{|c|r|r|r|r||r|r|r|r|}
\hline
Algorithm &\DR &$\epsilon_{\rm b}$ & $\rm {R_j}$&$\rm w_{cut}$ &\DR &$\epsilon_{\rm b}$ & $\rm {R_j}$&$\rm w_{cut}$\\
&& && &&&&\\
\hline
 SV2&0.7& 49\%  & 71& 1& 0.4 & 50\% &58&1\\
\hline
 3D&0.7& 55\%  & 27& 1&0.4&54\% &26&1\\
\hline
\hline 
 SV2&0.7& 46\%  & 219& 2&0.4&46\%&200& 2\\
\hline
 3D&0.7& 49\%  & 52& 2&0.4&49\% & 50& 2\\
\hline
\end{tabular}
\end{center}
\caption{ b-tag performance. The table shows, for the two algorithms used, the jet opening angle \DR, the efficiency on b-jet 
 identification $\epsilon_{\rm b}$, the rejection of light jets  $\rm {R_j}$ and the value of $\rm w_{cut}$ (see text).} 
\label{tab:b_tag }
\end{table}

As results of this study, it is possible to conclude that:
\begin{itemize} 
\item{} The SV2 algorithm, which uses  more information on b decays products, provides a higher rejection rate with little loss
in efficiency. 
\item{} No relevant improvement is obtained by shrinking the jet cone opening angle, $\DR$,
from 0.7 to 0.4. Therefore the ATHENA (v. 10.0.1) default value, $\DR$=0.7, will be used in the following analysis.
\item{}
The cut value $\rm w_{cut}$ = 1 turns out to be  a good compromise with a b-jet identification efficiency of $\approx$ 50\%
and a rejection of light jets by a factor of $\approx$ 70.
\end{itemize} 
In case of high luminosity, possible track pileup doesn't have a significant influence on b-tagging
efficiency for this class of events.  Their topology with two muons emerging from the primary vertex facilitates
the selection of the correct primary vertex out of many interaction vertices.

\subsection {Method for ${\boldsymbol \BBZ }$ background subtraction}\label{sec:Prelim3}
The method proposed in \cite{Gentile06} exploits the two following points (at the level of particle generation):\\
a) the rate of h/A $\ra \ee $ is expected to be suppressed with respect
   to the signal h/A $\ra \mm $ by a factor  $\Big(\frac{\Mmu}{\Me}\Big)^2$, \\
b) the rate of the background  $\bb \Zo \ra \bb \mm $  is equal to the rate
   of $\bb \Zo \ra \bb \ee$ because of the production diagrams which are the same,
   and of the lepton coupling universality in the Z decay. \\
In this context the  associated \Zo\ production and decay in the channel $\bb\Zo\ra \mm $ has been studied using a control 
sample of $\bb\Zo\ra \ee $ events. The effect of {\it inner bremsstrahlung} (IB) radiation has been investigated and corrected 
for; the impact in the event reconstruction is not large. The ratio of the number of reconstructed events from the two samples 
in the region of mass  higher than \MZ, interesting for new physics search, is stable and implies correction factors close to one
\cite{Gentile06} .  
\par
As a result, barring corrections for different {\it inner bremsstrahlung} and  detector response, the number of $\bb\Zo \ra \bb\ee$ 
gives directly the number of background events $\bb\Zo \ra \bb \mm$. Details of the method are given in Appendix \ref{sec:App_1}. 

\subsection{Significance of the search}\label{sec:Signif}
Statistical methods used in Higgs boson searches are discussed in Ref. \cite{Cranmer:2003kq}. Here
the  significance of a search is given using \SIGN\ as a statistical estimator, where S is the number of signal events 
(h/A or h/A/H), and B the number of background events. Discovery means that the signal is larger than 5 times 
the background statistical error ($\SIGN \geq 5 $). The probability of a background fluctuation of this size is less than 
$\approx 2.87 \cdot 10^{-7}$.  A search resulting in $\SIGN ~\geq  3 $ is interpreted as an indication of new physics.

\section{Monte Carlo data analysis}\label{sec:Cut}
\subsection{${\boldsymbol \BBA ,~\boldsymbol \BBh ,~\boldsymbol \BBH  }$}\label{sec:Cut1}
The signature of the  h/A  channel is a pair of well isolated high-energy muons with opposite charge  and two hadronic jets 
containing b quarks. The invariant mass of the reconstructed muons is supposed to originate from a h or A boson and must be 
compatible, within the mass resolution, with the corresponding mass, \Mh\ or \MA. 
\par
No simulation of the trigger is implemented; however two muons of  transverse momentum  above trigger experimental 
threshold are required and the detector acceptance considered is $\vert \eta\vert \leq 2.5$. The  event selection is 
divided in three steps: 
{\it preselection}, {\it tt veto selection}, {\it final selection}.

\par
The {\it preselection}, cuts 1-2-3 in Table \ref{tab:tablecuts }, requires at least two well identified opposite
charge muons with $\PTmu ~\geq~ 10 ~ \GeV$ in the  pseudo-rapidity  range $\vert ~\Etamu \vert \leq~2.5 $ (cut 1), thus 
satisfying the trigger requirement. The presence 
of a jet pair with $\PTjet \geq 10~\GeV$ and $\vert ~\Etajet \vert \leq~2.5 $  is as well demanded, without any b-identification 
requirement (cut 2).
 
\noindent A further requirement (cut 3) is that at least one of these jets be identified as originating from a b-quark with 
$\PTjet > 15$ \GeV, that is the energy lower limit for a reliable b-jet identification \cite{Kostyukhin03} (Sec. \ref{sec:btag}).
These cuts are designed to select events fulfilling the minimum conditions to be analyzed later. The events excluded are in any 
case not suitable to undergo any further analysis.

\par 
The {\it tt veto selection}~ is designed to suppress this background, characterized  by a large missing transverse energy due to 
the presence of neutrinos. \ETmiss\ is  required to be less than 45 GeV (cut 4).
Other cuts (cut 5-6) are applied on the first and second most large muon transverse momentum, \PTmumost\ and \PTmuleast, 
and (cut 7-8) on the first and second most large jet transverse momentum, \PTjetmost\  and  \PTjetleast\ . 
At least one of these jets was previously identified as originating from a b-quark (cut 3).

\par 
The  {\it final selection} criteria are designed mainly to disentangle the signal events from the irreducible \Zo\ background.
\noindent The main requirement is that the \mm\ invariant mass has to lie inside a window around h/A mass, determined by the 
natural width of the bosons and by the experimental resolution of the \mm\ invariant mass, Sec. \ref {sec:Prelim}.

\par  
The muons originating from b decays in bbbb events, with a cross section of $\approx 500$ pb \cite{Mangano:2002ea}, can mimic 
the signal events. To avoid this possible contamination a low hadronic activity near both muons is required. The isolation 
criteria demands that the sum of the charged track momenta in a cone ($\DR <0.2$) around  the muon direction be less than 5 \GeV.
These  selection cuts are summarized in Table \ref{tab:tablecuts }.

\begin{table}[h]
\begin{center}
\begin{tabular}{|r|c|}
\hline
Cut Number & Variable\\
\hline
1&\1cut \\
\hline
2&\2cut  \\
\hline
3&\3cut \\
\hline
4&\8cut \\
\hline
5&\4cut\\
\hline
6&\5cut\\
\hline
7&\7cut \\
\hline
8&\6cut \\
\hline
9&\9cut \\
\hline
10&\cutdieci\\
\hline
\end{tabular}
\end{center}
\caption{ Selection criteria for h/A $ \ra \mm$ with two b-quarks final state.  }
\label{tab:tablecuts }
\end{table}

\par A key point of the selection is the determination of the mass window used (cut 9). Its value is determined as a function of  
\Gh\ (\GA ), the total width of the h (A) Higgs boson, and the experimental mass resolution $\sigma_{\rm m}$ = 2.6\% 
(Sec.~\ref{sec:Zmm}). The mass window is centered on the h (A) mass corrected for the bias k in the muon reconstruction, 
$m_{\h,\rm{A}}^{\rm corr} =  m_{\h,\rm{A}} - \rm{k}$  with k= $-820$ \MeV\ (Sec.~\ref{sec:Zmm}). Its limits,
$\xi_{\h,\rm A}^{\pm}$, are defined by:  
\begin{equation}
\xi_{\h,\rm A}^{\pm} = m_{\h,\rm{A}}^{\rm corr} \pm \rm{f} \cdot 
\Bigg(\bigg(\frac {\Gamma_{\h,\rm{A}}}{2.36}\bigg)^2 + \sigma^2_{\rm m}\Bigg)^\frac{1}{2},
\label{eq:window}
\end{equation}
\noindent 
where f is the standard deviation factor corresponding to a chosen probability. In our analysis f = 2, which corresponds to 
including 97.7\% of the signal. The event is selected  if the \mm\ invariant mass is inside the window limits (\ref{eq:window})
either for the A or the h mass. 
 
For the \BBH\ channel the procedure is the same as for the h/A search, but for the selection window. The window limits are obtained 
from Eq.~\ref{eq:window} with \MH\ (the H mass) replacing  $m_{\h,\rm{A}}$ in $m_{\h,\rm{A}}^{\rm corr}$.

\subsection{Reference point}\label{sec:RefPoint} 

The analysis procedure is first discussed for the reference point  \MA\ = 110.31 \GeV, \tanb\ = 45 (\Mh\ = 110.00 \GeV, \MH\ = 
127.46 \GeV ). For a given MSSM neutral Higgs boson (h, A, H) at this point, Table \ref{tab:reference point} shows the production 
cross section times the \mm\ decay branching ratio, $\sigma_{\bb\mm }$, the expected number of signal events at $\IL = 300$ 
\fbinv, $\rm N^{\rm exp_{300}}$, the number of Monte Carlo generated events $\rm N^{\rm MC}$ and their weight ${\rm w}$.
The weight of the signal events is close to one as it is for all (\MA, \tanb ) points\cite{Gentile07}. The background weights 
(Table \ref{tab:background}) are also close to one.\\
We shall justify the cuts by showing examples of distributions obtained for the \BBh\ signal and the \BBZ\ background, and refer 
otherwise to \cite{Gentile07} either for variables different from the ones displayed or for a different signal (\BBA\ or \BBH ) or 
for a different background (\BBTT\ or \ZZ ). In all cases the reconstructed distributions without cuts compare to the generated 
distributions as expected.       

\begin{table}[h]
\begin{center}
\begin{tabular}{|c|r|r|r|r|}
\hline
process& $\sigma_{\bb\mm}$ & $\rm N^{\rm exp_{300}}$ & $\rm N^{\rm MC} $ &   ${\rm w}$ \\
(signal) &[pb]~~~~~~& & & \\

\hline
\BBh&  0.245      & 73500& 76517 & 0.96 \\
\hline
\BBA&  0.2433     & 72900& 74965 & 0.97 \\
\hline
\BBH&  0.001619   &   486&   600 & 0.88 \\
\hline
\end{tabular}
\end{center}

\caption{h, A, H production at the reference point \tanb\ = 45, \MA\ = 110.31 \GeV (\Mh\ = 110.00 \GeV, \MH\ = 127.46 \GeV ): 
         process, cross section times \mm\ decay branching ratio, $\sigma_{\bb\mm}$, number of events expected for 
         $ \IL =  300$ \fbinv, ${\rm N^{\rm exp_{300}}}$, number of Monte Carlo generated events, ${\rm N^{\rm MC}}$, their 
         weight w.
         }

\label{tab:reference point}
\end{table}
\par 
The distributions of the number of muons \Nmu\  and of the muon transverse momentum \PTmu\ in the processes \BBh\ and \BBZ\ are 
displayed in Fig.~\ref{fig:Nocut_Nmmuon_text} for the generated (hatched) and reconstructed (full color) events without any cut 
applied . 

\begin{figure}
\begin{tabular}{cc}
\hspace{-7mm}
\includegraphics*[width=0.5\textwidth]{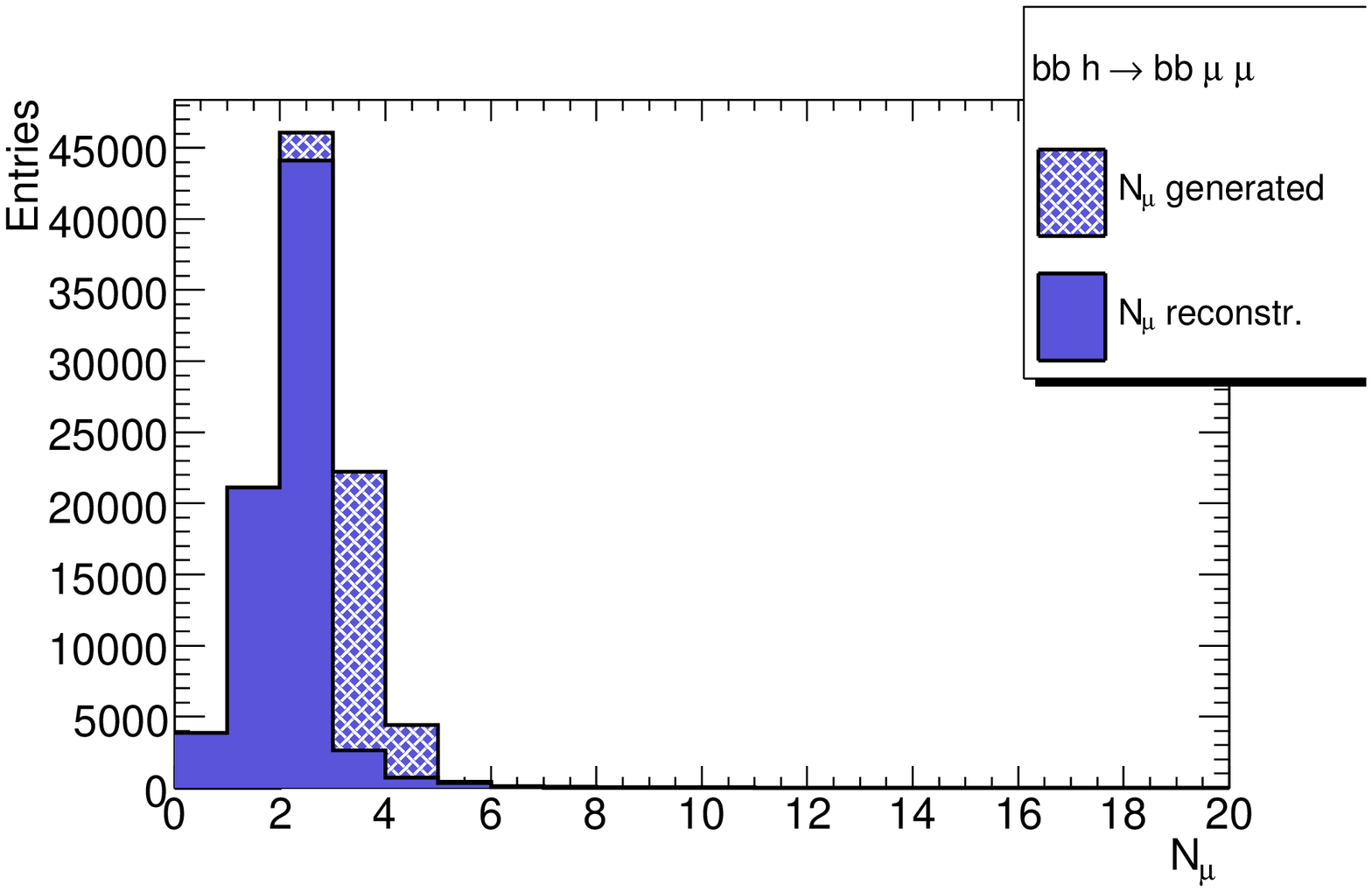} &
\hspace{-0mm}
\includegraphics*[width=0.5\textwidth]{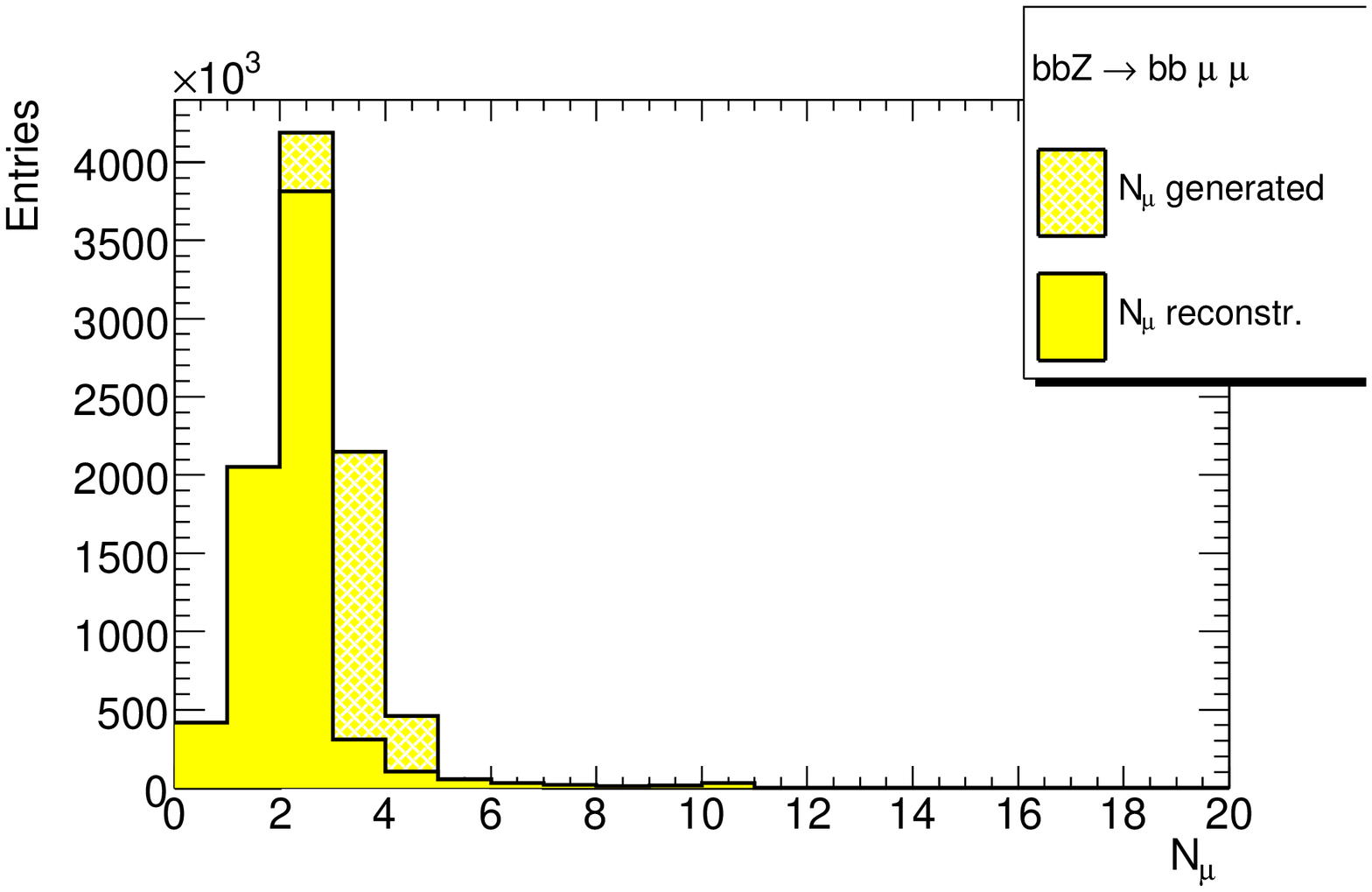} \\
\hspace{-7mm}
\includegraphics*[width=0.5\textwidth]{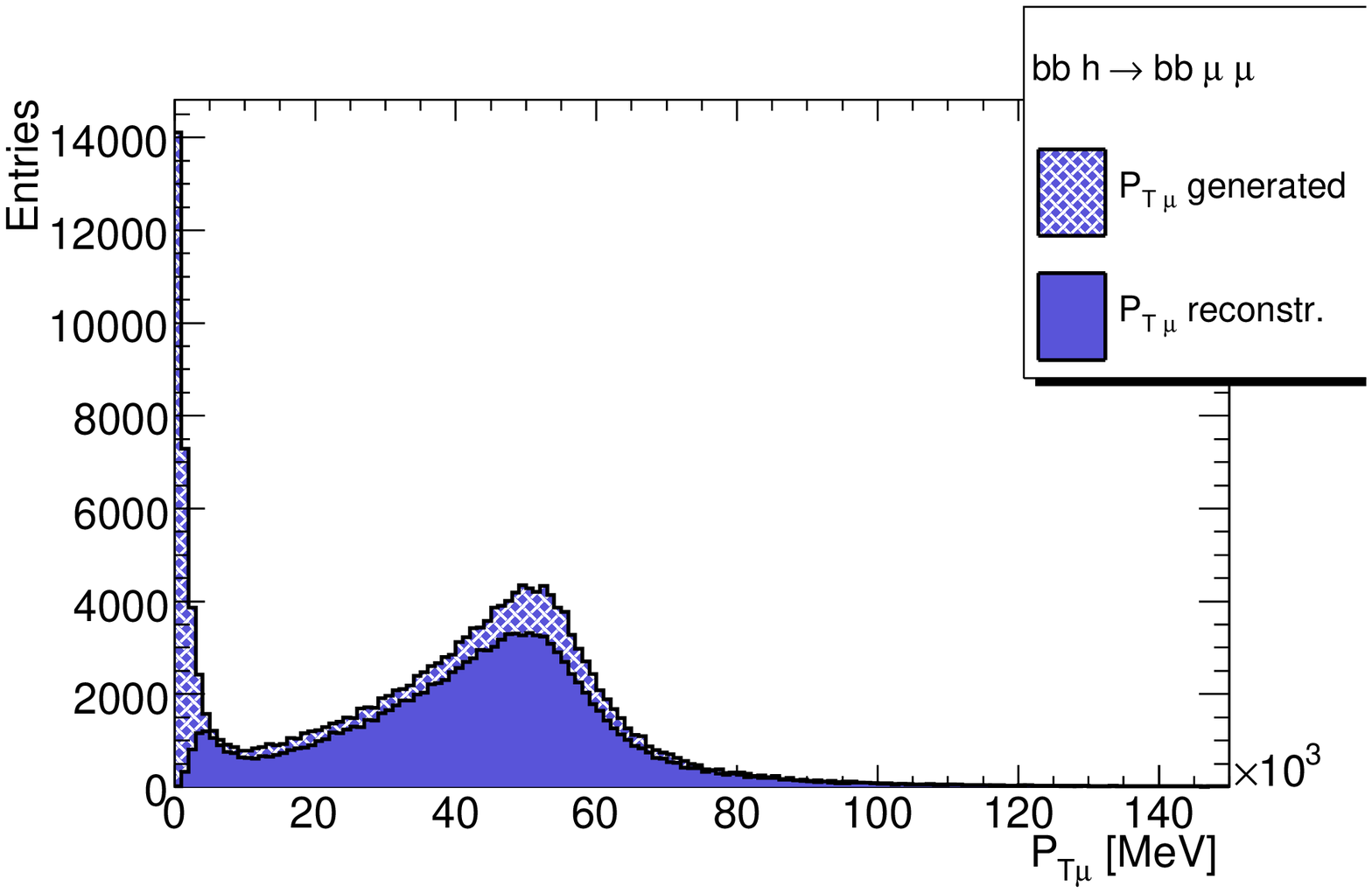} &
\hspace{-0mm}
\includegraphics*[width=0.5\textwidth]{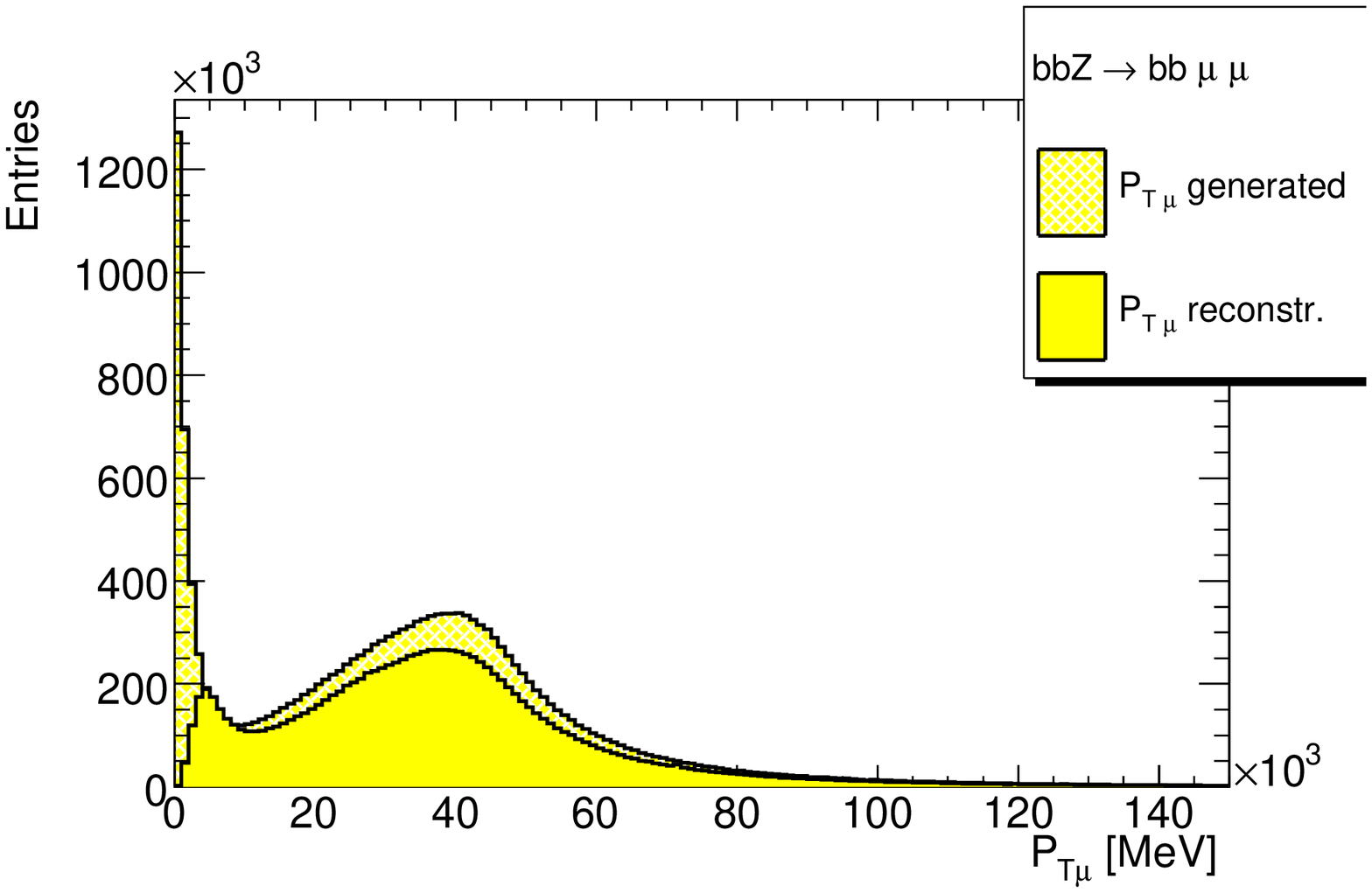} \\
\end{tabular}
   \caption[]{\label{fig:Nocut_Nmmuon_text}
           {\bf Without cut.}
           Distributions of the number of muons \Nmu\ (top) and of the transverse momentum \PTmu\ (bottom) for generated (hatched) 
           and reconstructed (full color) events are plotted without cuts applied: \\ 
           a) \BBh\  events at the reference point (\tanb\ =  45, \Mh\ = 110.00 \GeV )(left, dark blue);  
           b) \BBZ\ events (right, yellow). All distributions are normalized at \IL\ = 300 \fbinv.
           Entries are per  bin width  of 1 (top plots) and of $ 10^3$ MeV (bottom plots).
           }
\end{figure}
\par 
The  {\it preselection}  requirement on muon pair (cut 1, Table \ref {tab:tablecuts }) reduces approximately to same fraction of
signal ($\sim$ 62\%) and background (\Zo $\sim$ 59\%, \tt\  $\sim$ 67\%) events. The requirement on jet pair (cut 2) leaves almost 
untouched \tt\  ($\sim$ 63\%) reducing \Zo\  to  $\sim$ 54\%,  because of the more energetic \PTjet~ spectra of the top decay 
\cite{Gentile07}.
Unfortunately for the same reason this cut reduces the signal sample significantly to $\sim$ 30\% , 
see Table \ref {tab:tables_presel}.
\par  
The mean  jet  transverse   momentum  $<\PTjet>$ is $\sim$  26 \GeV~ for the h boson,  and $\sim$ 36  \GeV\ for 
\Zo~ decays, see Fig.~\ref{fig:2cut_PTjet_text} \cite{Gentile07}. Then, as stated in Sec.~\ref{sec:Background}, 
the requirement that two b-jets be identified will not  enhance the signal over this specific background. This requirement  will 
suppress  more the signal than  the \Zo\ background, since the two b-jets of this background  events are usually more energetic
than those accompanying the signal.  According  to the study in Sec. \ref{sec:btag}, only one jet is required to be tagged as a
b-jet.
\par 
After  the three preselection cuts (1-2-3) the signal sample of h or A is  reduced  to $\sim$ 10\%. 
The  backgrounds are reduced to $\sim$ 17\%  for the \Zo, $\sim$ 53\% for \tt, and  $\sim$ 28\% for \Zo\Zo.

\par 
The {\it tt veto selection} is designed to exploit features of t decays. 
The \tt~ sample is characterized by a transverse missing energy  \ETmiss, which, due to neutrinos, is  larger than  in the h/A sample, 
as shown in Fig.~\ref{fig:3cut_ETmiss_text}.
The \ETmiss\ cut (cut 4, Table \ref {tab:tablecuts }) is extremely effective, reducing \tt~ to  $\sim$ 14 \%  of the original sample 
while keeping the signal  at $\sim$ 9 \% almost unchanged and the  \Zo\ background at $\sim$ 14 \%. The two cuts (5-6), related to
$\PTmu$ distributions, reduce the \tt\ original sample to $\sim$ 8 \%, but are little effective on the \Zo\ sample (to $\sim$ 10\%)
and the signal samples of h and A (to $\sim$ 8 \%). 
A further reduction of the \tt~ background to 1.4 \% is obtained by applying cuts (7-8) on  $\PTjet$. At this stage, the  \Zo~ sample 
is reduced to  $\sim$ 5.8 \% while the h signal as well as the A signal is reduced to 6.0 \%, see Fig.~\ref{fig:6cut_Minv_text}.

\begin{figure}
\begin{tabular}{cc}
\hspace{-7mm}
\includegraphics*[width=0.5\textwidth]{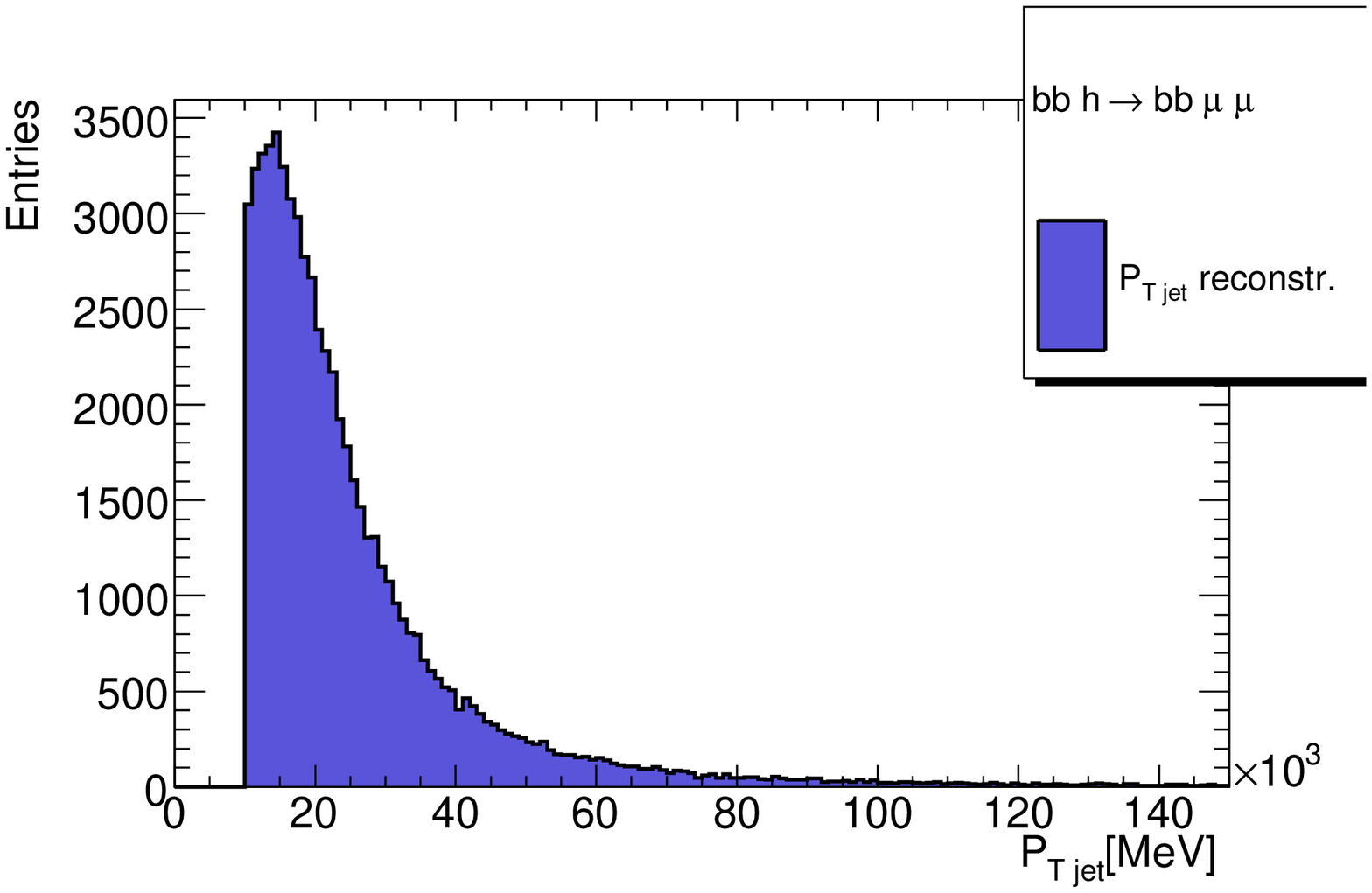} &
\hspace{-0mm}
\includegraphics*[width=0.5\textwidth]{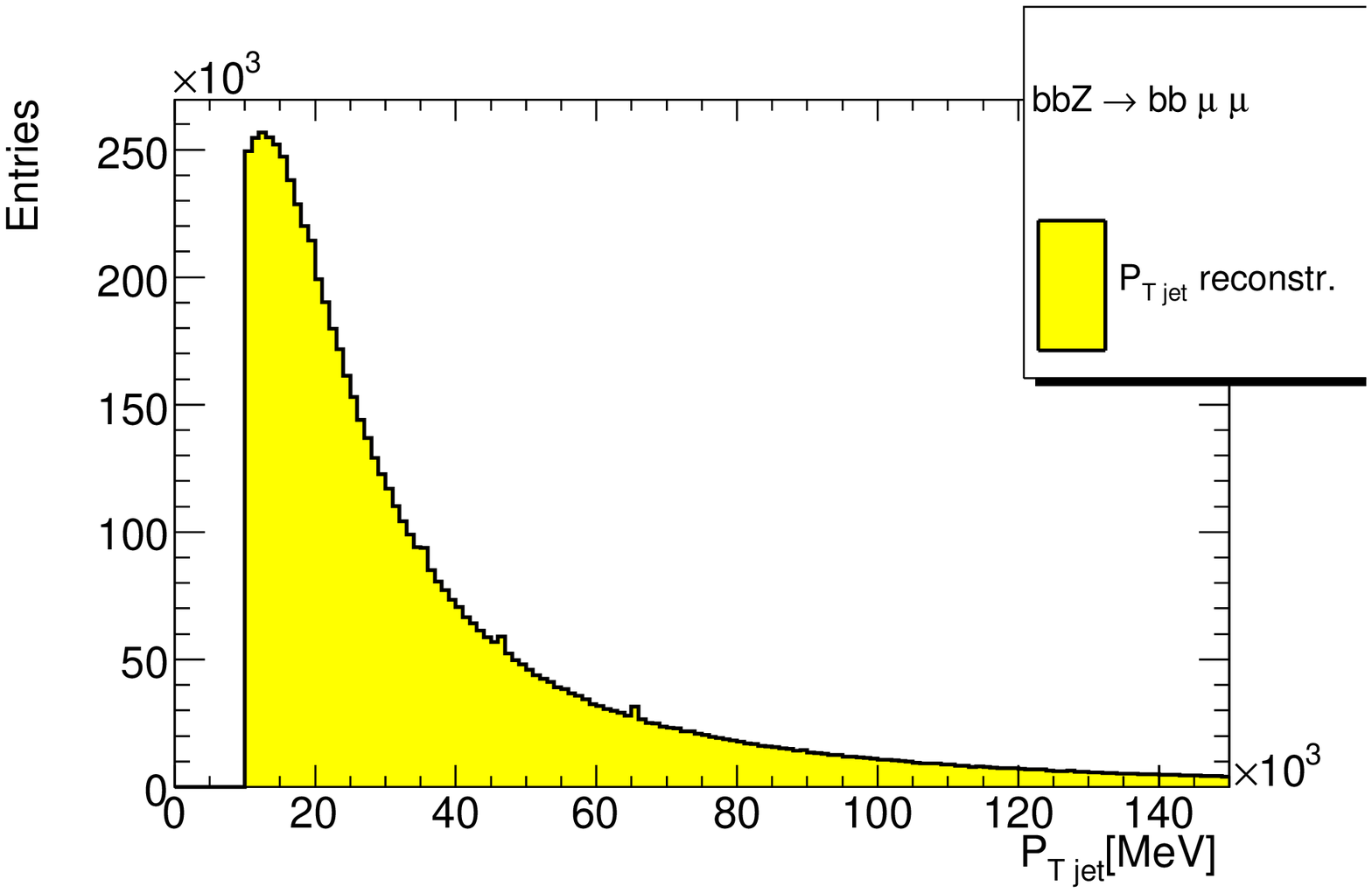} \\
\end{tabular}
    \caption[]{\label{fig:2cut_PTjet_text}
            {\bf After 1-2 cuts.}
            Distributions of the reconstructed jet transverse momentum  \PTjet\ are plotted after cuts 1 and 2 
            (Table \ref{tab:tablecuts }) for : \\ 
            a) \BBh\ events at the reference point (\tanb\  = 45, \Mh\  = 110.00 \GeV), (left, dark blue);
            b) \BBZ\  events (right, yellow). All distributions are normalized at \IL\ = 300 \fbinv.
            Entries are per bin width  of $ 10^3$ MeV.
           } 
\end{figure}


\begin{figure}
\begin{tabular}{cc}
\hspace{-7mm}
\includegraphics*[width=0.5\textwidth]{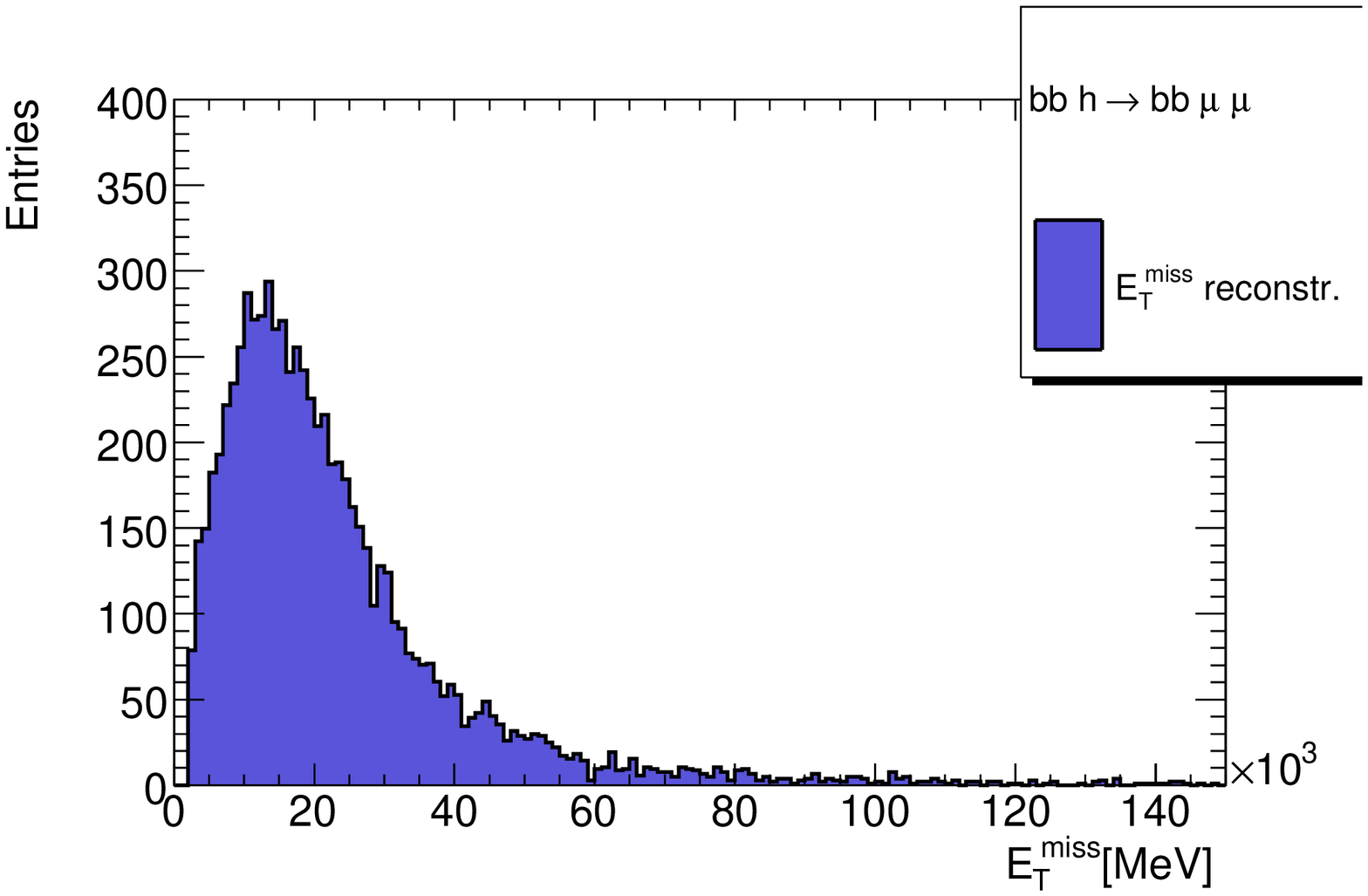} &
\hspace{-0mm}
\includegraphics*[width=0.5\textwidth]{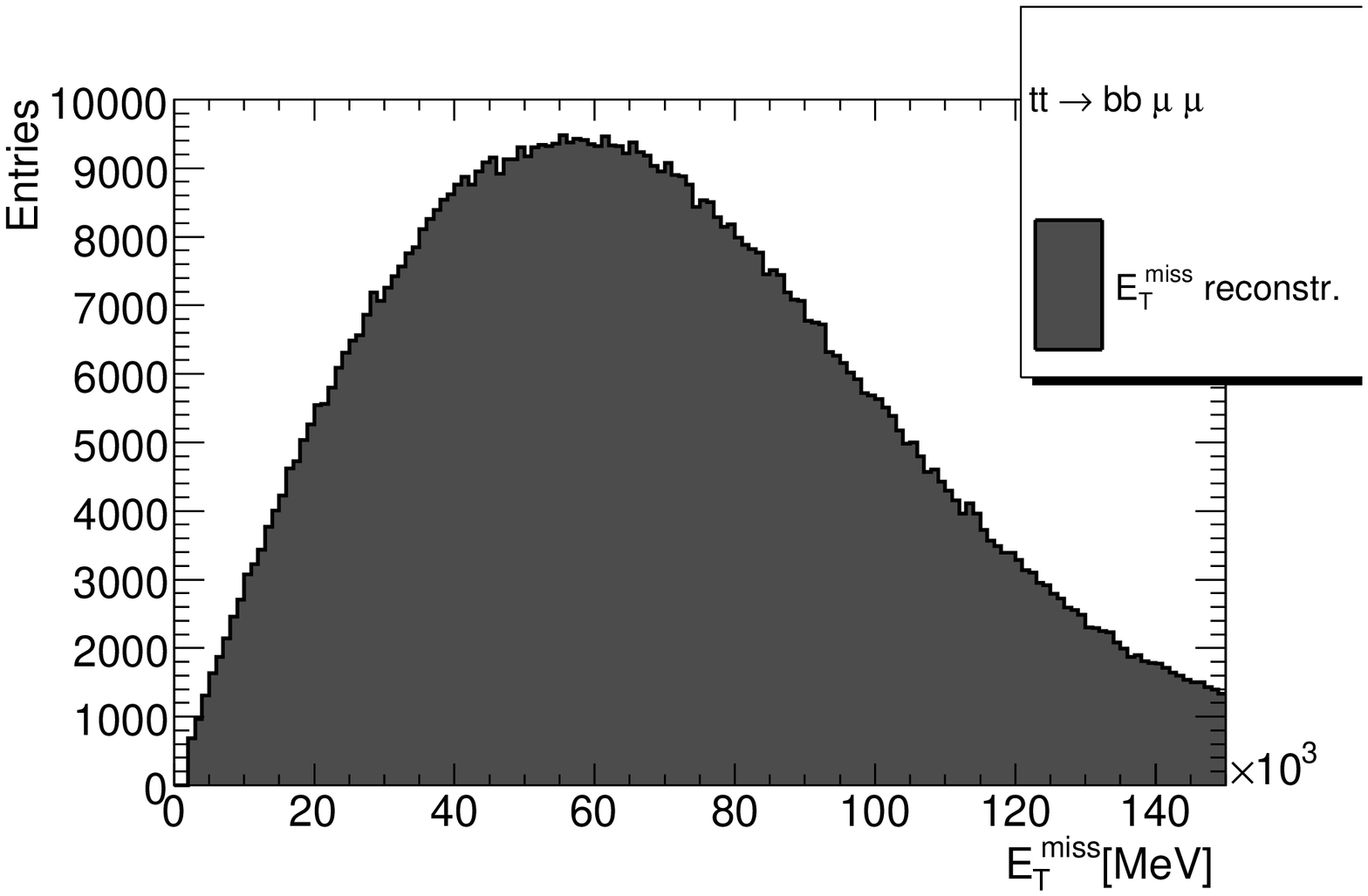} \\
\end{tabular}
   \caption[]{\label{fig:3cut_ETmiss_text}
            {\bf After 1-2-3 cuts.}
            Distributions of the reconstructed transverse missing energy \ETmiss\  are plotted after cuts 1, 2  and 3 
            (Table \ref{tab:tablecuts }) for : \\ 
            a) \BBh\ events at the reference point (\tanb\  = 45, \Mh~ = 110.00 \GeV), (left, dark blue); 
            b) \BBTT\ events (right, gray). All distributions are normalized at \IL\ = 300 \fbinv.
             Entries are per bin width  of $ 10^3$ MeV.
            }
\end{figure}
\par 
The effect of cuts (1-8) is summarized in
Fig. \ref{fig:pre_Minv_final} which shows the distributions of the reconstructed \mm\ invariant mass  for signal
(h and A) and (weighted) background (Z, \tt\ and ZZ added up) events. The h/A signal (light blue) is clearly  
visible on top of the remaining background events (\Zo, \tt\ and \Zo \Zo\ added up, dark brown). 
\par
To evaluate the significance of observation 
of a signal, the mass window (Eq. \ref{eq:window}) is then selected, {\it final selection} (cut 9, Table \ref {tab:tablecuts }). 
At the reference point (\MA\ = 110.31 \GeV, \GA\ = 4.28 \GeV\ and \Mh\ = 110.00 \GeV, \Gh\ = 4.20 \GeV ), Eq. \ref{eq:window} 
implies to look for the A signal in a window, $\xi_{\rm A}$, of 3.394 \GeV\ around a
corrected mass value, $\MA^{\rm corr}$, of 109.490 \GeV , and for the h signal in a window, $\xi_{\h}$, of 3.369 \GeV\ around  
$\Mh^{\rm corr}$ = 109.180 \GeV\ (see \cite{Gentile07} for the others values of \tanb, \MA\ and \Mh ). As for the 
probability requirement f = 2,  the windows $\xi_{\h}$ and $\xi_{\rm A}$ are increased by a factor 2 to obtain the effective 
selection window used in Table \ref {tab:tables_finalsel}.
\par 
This selection reduces drastically the background samples, the \Zo\ to less than 0.13\%, the \tt\ to 0.17 \% and the \Zo\Zo\ to a 
negligible 0.24 \%. The event signal surviving in both A and h channels is $\approx$ 4.2 \%.
\par  
The last reduction (cut 10) on muon isolation does not change significantly the previous results. Indeed this cut is expected
to be effective in the data against the heavy flavour QCD background, and cannot be tested properly in Monte Carlo.


\begin{figure}
\begin{tabular}{cc}
\hspace{-7mm}
\includegraphics*[width=0.5\textwidth]{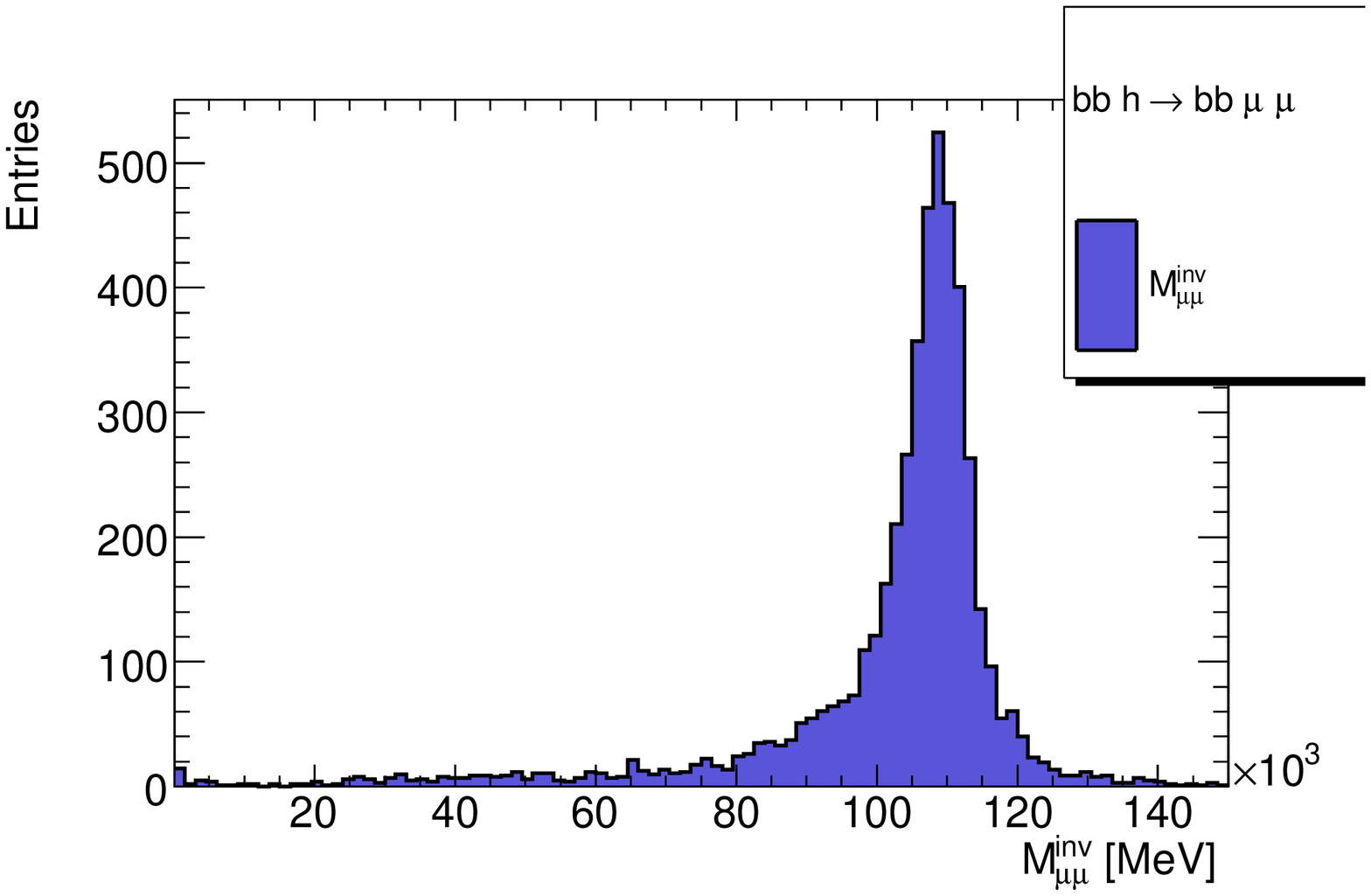} &
\hspace{-0mm}
\includegraphics*[width=0.5\textwidth]{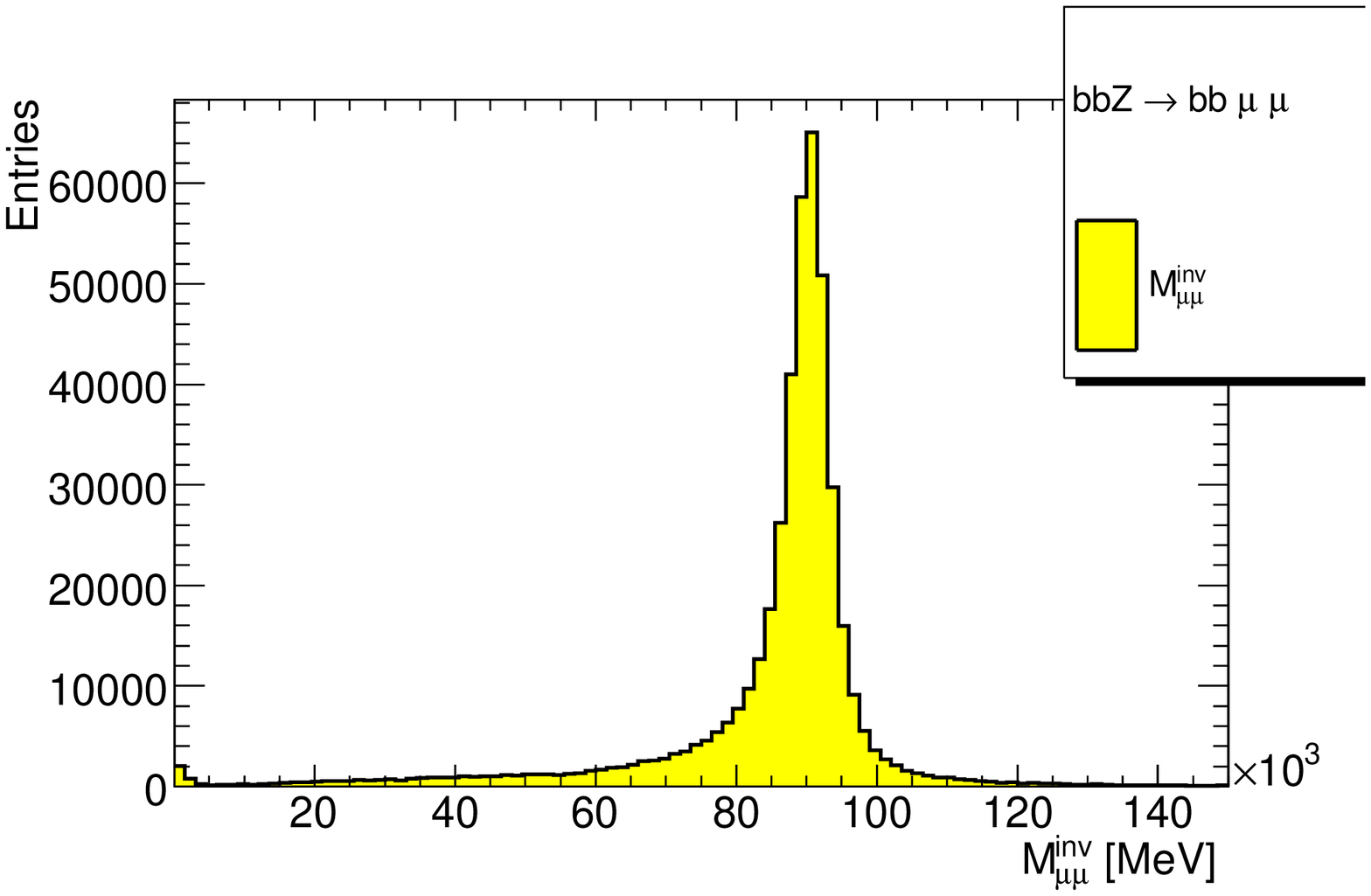} \\
\end{tabular}
    \caption[]{\label{fig:6cut_Minv_text}
              {\bf After 1-8 cuts.}
              Distributions of the reconstructed \mm\ invariant mass \Minv\ are  plotted after cuts 1 -- 8 
              (Table \ref{tab:tablecuts })  for: \\ 
              a) \BBh\ events at the reference point (\tanb\  = 45, \Mh\ = 110.00 \GeV ) (left, dark blue); 
              b) \BBZ\ events (right, yellow). All distributions are normalized at \IL\ = 300 \fbinv.
              Entries are per bin width  of $1.5 \cdot 10^3$ MeV.
               }

\end{figure}


\begin{table}[h]
\begin{center}
\begin{tabular}{|c|r|r|r|r|r|}
\hline
Process& $\rm N^{\rm exp_{300}} $ & $\rm N^{\rm MC}$& $\rm {N_{\rm 1 cut}}$ & $\rm {N_{\rm 2 cut}}$ & $\rm {N_{\rm 3 cut}}$ \\
 &&& & &\\

\hline
\BBh & 73500& 76517 & 47099 & 22969&7907\\
\hline
\BBA& 72900&74965 & 46247& 22693&7919\\
\hline
\BBH & 486 &600&394 &222&95\\
\hline
\BBZ  & 6836700 &3314000& 1945387 & 1178864&579751\\
\hline
\BBTT& 1713420&  1806437 &1217706 & 1137780&953928\\
\hline
\ZZ  & 33819& 97244& 46487 & 38600&27744\\
\hline
\end{tabular}
\end{center}
\caption{
        {\it Preselection.} Signal/Background process, number of events expected at $\IL = 300~ \fbinv$, 
        $\rm N^{\rm exp_{300}}$, number of Monte Carlo events generated, $\rm N^{\rm MC}$, and after three preselection steps, 
        $\rm {N_{\rm 1 cut}}$,~$\rm {N_{\rm 2 cut}}$ and $\rm {N_{\rm 3 cut}}$ (Table \ref{tab:tablecuts }). The signal is 
        evaluated at the reference point (\tanb =45~, \MA\ = 110.31 \GeV, \Mh\ = 110.00 \GeV ). Cuts 1 to 3 are described in 
        Table \ref  {tab:tablecuts }}
\label{tab:tables_presel}
\end{table}


\begin{table}[h]
\begin{center}
\begin{tabular}{|c|r|r|r|r|r|r|}
\hline
Process& $\rm {N_{\rm 3 cut}}$ & $\rm {N_{\rm 4 cut}}$& $\rm {N_{\rm 5 cut}}$  & $\rm {N_{\rm 6 cut}}$
                               & $\rm {N_{\rm 7 cut}}$ & $\rm {N_{\rm 8 cut}}$\\
                               &&& & & &\\

\hline
\BBh & 7907&7099 & 6234&5575 &5085& 4636\\
\hline
\BBA& 7919& 7096& 6183 &5575 &5118&4595\\
\hline
\BBH & 95& 78& 66 &54 &50&44\\
\hline
\BBZ  & 579751&478106 &397030 &334609 &226414&193334\\
\hline
\BBTT& 953928&256662&190672 & 143746&41954&25567\\
\hline
\ZZ  & 27744&24516& 19102 & 15957&10926&8574\\
\hline
\end{tabular}
\end{center}
\caption{{\it tt veto selection.} 
        Signal/Background process, number of Monte Carlo generated events after the preselection cut 3 and cuts 4 to 8.
        The signal is evaluated at the reference point (\tanb =45~, \MA\ = 110.31 \GeV, \Mh\ = 110.00 \GeV ). Cuts 3 to 8 
        are described in Table \ref{tab:tablecuts }. 
         }
\label{tab:tables_ttsel}
\end{table}


\begin{table}[h]
\begin{center}
\begin{tabular}{|c|r|r|r|}
\hline
Process& $\rm {N_{\rm 8 cut}}$ & $\rm {N_{\rm 9 cut}}$ & $\rm {N_{\rm 10 cut}}$  \\
 &&& \\

\hline
\BBh &4636 & 3212&3165\\
\hline
\BBA&4595 &3151&3110\\
\hline
\BBH &44&&\\
\hline
\BBZ& 193334&4496&4206 \\
\hline
\BBTT& 25567&3148&2969\\
\hline
\ZZ  & 8574 &260&234\\
\hline
\end{tabular}
\end{center}
\caption{
        {\it Final selection.} 
        Signal/Background process, number of Monte Carlo generated events, after cut 8  and successive cuts 9 and 10. 
        The signal is evaluated at the reference point (\tanb\ = 45, \MA\ = 110.31 \GeV, \Mh\ = 110.00 \GeV ). Cuts 8 to 
        10 are described in Table \ref{tab:tablecuts } (see text for cut 9 window). }
\label{tab:tables_finalsel}
\end{table}

\par
We calculate that the significance \SIGN\  at  $\IL =  300$ \fbinv\ is $\approx$ 56, scaling down to $\approx$ 
18 at $\IL = 30$ \fbinv\ and $\approx$ 10 at $ \IL = 10$ \fbinv .
As discussed in Sec. \ref{sec:Background}, the process of Z production accompanied by two light jets, which has not been simulated,
can be an additional significant background. Its importance has been estimated from the relative cross-section and fake b-tagging
probability.  When this background is taken into account, we have estimated that the significance is lowered by $\approx$ 25\%. 
\par
We conclude that if  \Mh\ = 110.00 \GeV\ and consequently  \MA\ = 110.31 \GeV\ 
there is a high probability  for these bosons to be discovered at the beginning  of data taking.  


\begin{figure}[h]
\begin{center}
\includegraphics*[width=0.8\textwidth]{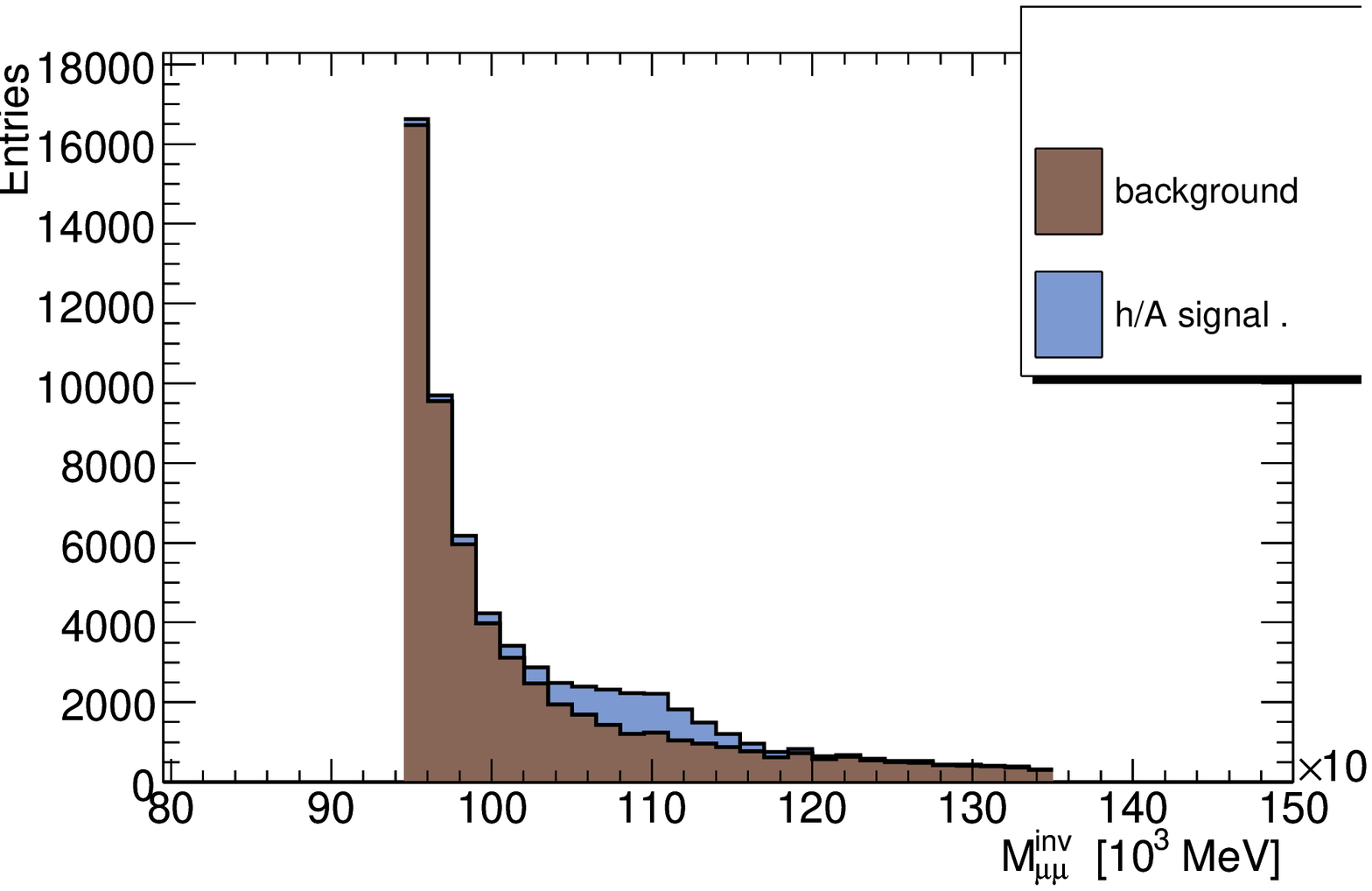} 
\end{center}
\caption{Distributions of the reconstructed \mm\ invariant mass, \Minv, for signal and backgrounds events, after the selection 
         cuts 1-8 (Table \ref  {tab:tablecuts }) at the reference point (\tanb\ = 45, \MA\ = 110.31 \GeV , \Mh\ = 110.00 \GeV ).
         The two distributions are normalized at \IL = 300 \fbinv. The h/A signal (light blue) emerge over the background 
         (\Zo, \tt\ and \Zo\Zo ) (dark brown). Entries are per  bin width  of $1.5 \cdot 10^3$ MeV.
         }
\label{fig:pre_Minv_final}
\end{figure}

\section{Search in the MSSM plane}\label{sec:Scan}
\subsection{${\boldsymbol \BBA }$, ${\boldsymbol \BBh }$ \\ }\label{sec:Scan_h}

A search for neutral Higgs bosons h/A has been performed in  the  (\MA, \tanb ) plane, varying  \tanb\ between 15 and 50, in steps 
of 5, and  \Mh\ between 95 \GeV\ and 125 \GeV, in steps of 2.5 GeV. 
For each point the statistics reached leads to  weight factors for h and A boson close to unity. 
\par  
The analysis described in Sec.~\ref{sec:Cut} has been repeated for all 104 simulated points and the results are reported in Ref.
\cite{Gentile07}.\\ 
The use of an  asymmetric selection window, centered at $m_{\h,\rm{A}}^{\rm corr}$ (see Sec. \ref{sec:Cut1}), has been 
tested for masses below 100 \GeV, with a view to exclude the \Zo\ events, however without obtaining  a significantly improved 
result ($\sim$ 10\%). A symmetric window as described in Sec.~\ref{sec:Cut1} was thus applied to all masses.\\
\par  
The search significance for the h/A  neutral boson is shown  as a function of \MA\ up to highest allowed value of \Mh , 
in Fig.~\ref {fig:Significance} for all scanned values of \tanb\ and three luminosity,  $\IL =  300$ \fbinv\ (\StrechA ), 
30 \fbinv\ (\StrenhA ) and 10 \fbinv\ (\SdiehA )\cite{Gentile07}. The values for the two lower luminosities were derived from 
the first one, which corresponds to the  highest statistics.. 
\par 
\noindent One should note that large h/A masses are penalized by a small cross section, thus implying a lower significance, while 
the masses near to \MZ\  suffer from the difficulty in disentangling the neutral Higgs boson signal from the \Zo\ background.
\par
The best mass range for an early discovery of h is between 100 and 120 \GeV\ at any given \tanb . If $\tanb > 30 $ 
a large range of masses is accessible to discovery even after the first year of data taking. More integrated luminosity, 
between $\approx$ 30 and 50 \fbinv , is needed for \tanb\  between 30 and 20. The discovery at \tanb\ = 15 demands 
a luminosity of $\approx$  150 \fbinv , making the exploration of this region possible only after a few years of 
data taking.
\par 
These considerations are summarized in Fig. \ref{fig:Luminosity_min}, where the minimum integrated luminosity \IL , demanded 
for a 5 $\sigma$ discovery of the  h/A neutral Higgs boson, is plotted as a function of \MA\ up to highest allowed value 
of \Mh , at given  \tanb\ values. With a $ \IL \approx  10$ \fbinv , corresponding to one year of data taking,
most of the masses are accessible if $\tanb >  30$. More integrated luminosity is needed  for \tanb~ = 20 and  \tanb~ = 15. 
Low masses need as well more luminosity in order to extract the evidence of a signal from the most copious \Zo\ background.
 
\begin{figure}
\begin{tabular}{cc}
\hspace{-7mm}
\includegraphics*[width=0.5\textwidth]{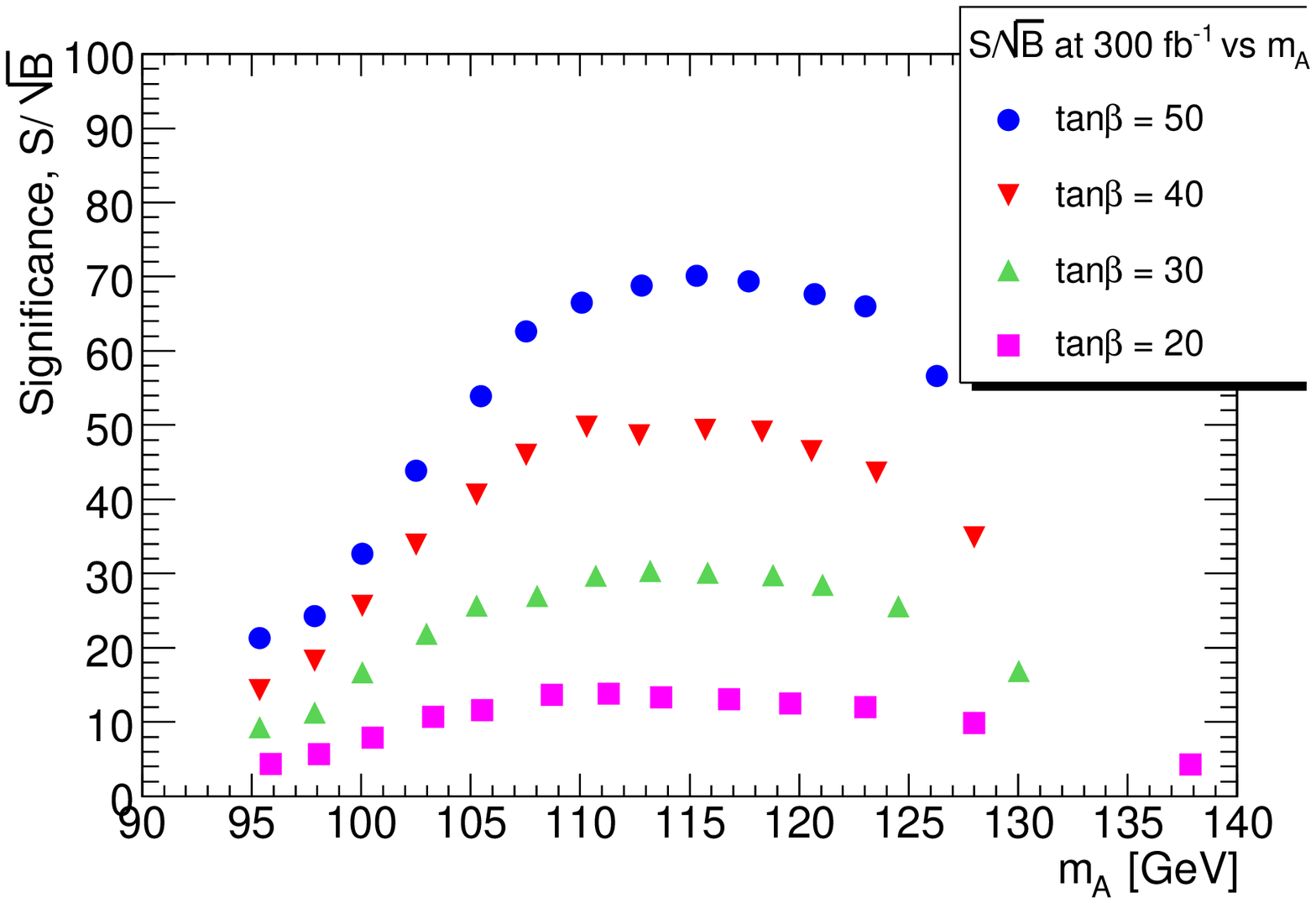} &
\hspace{-0mm}
\includegraphics*[width=0.5\textwidth]{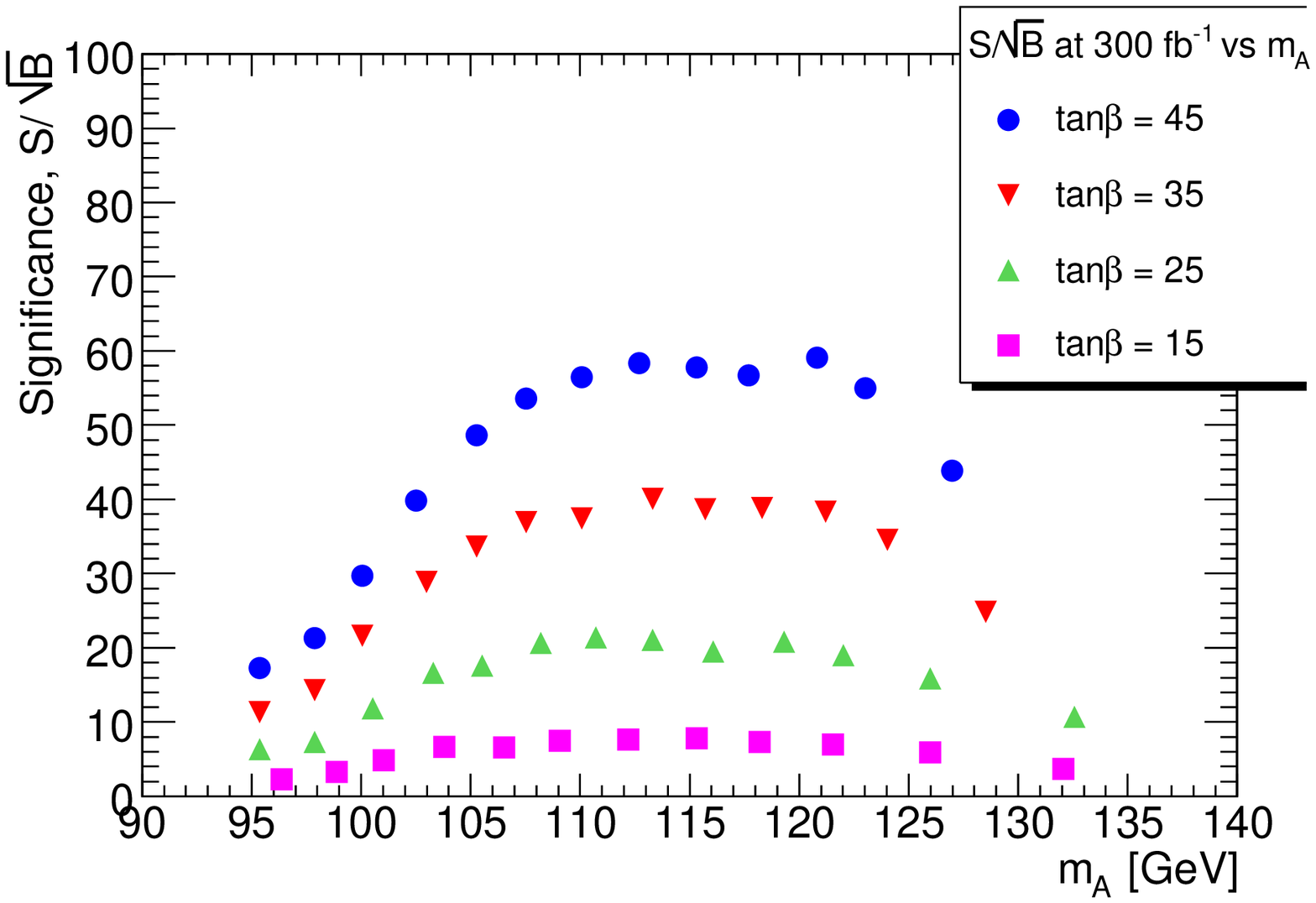} \\
\hspace{-7mm}
\includegraphics*[width=0.5\textwidth]{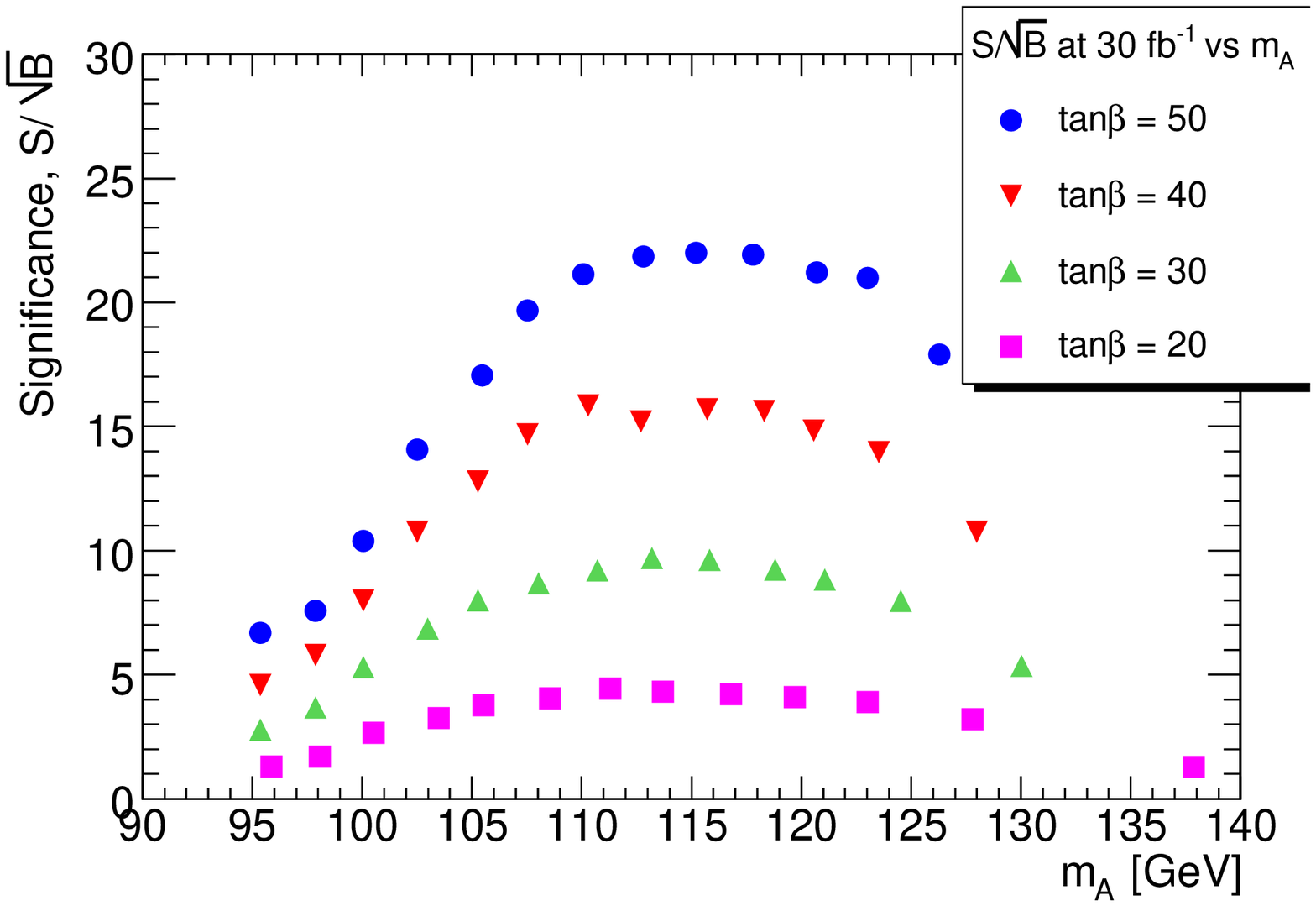} &
\hspace{-0mm}
\includegraphics*[width=0.5\textwidth]{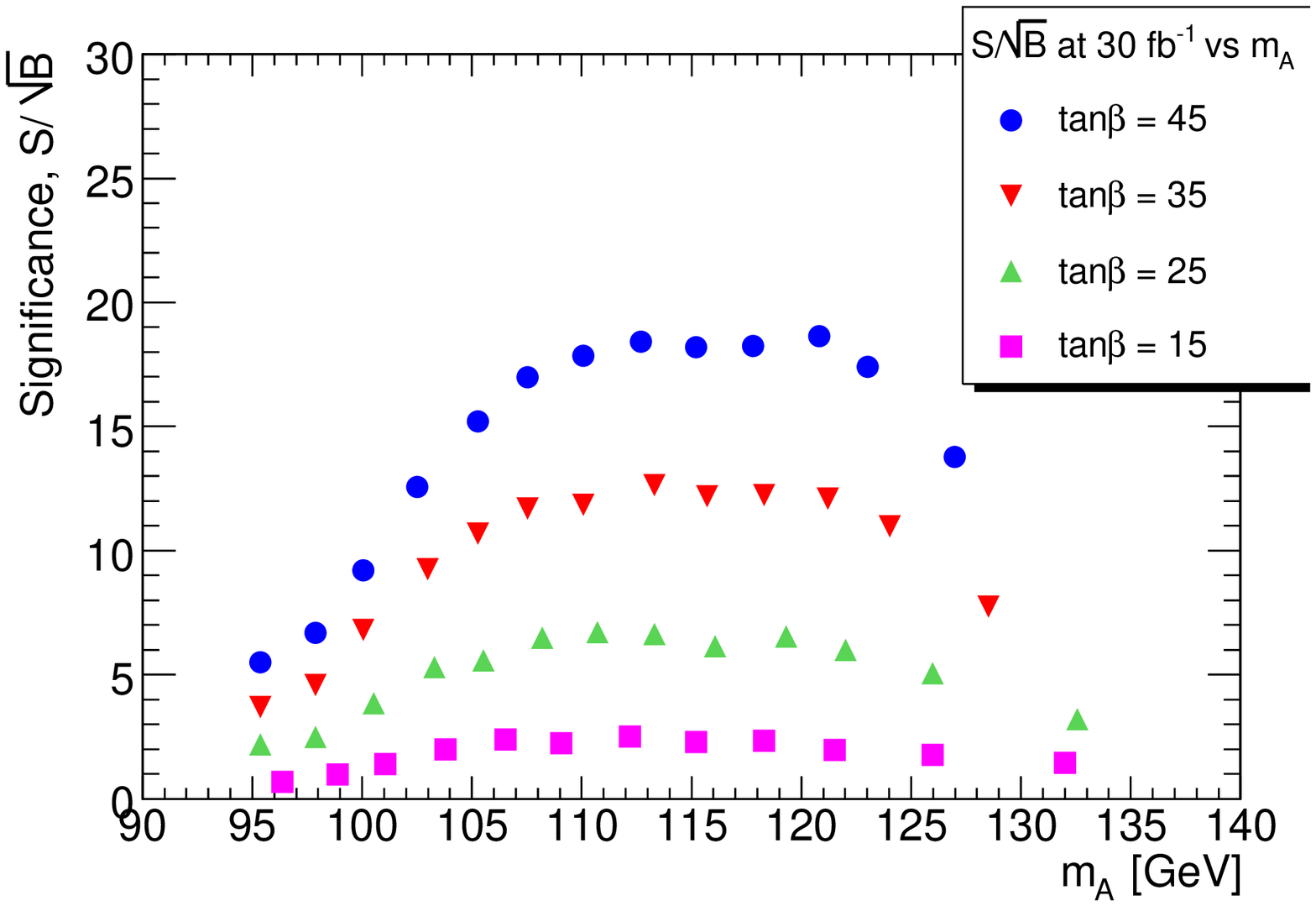} \\
\hspace{-7mm}
\includegraphics*[width=0.5\textwidth]{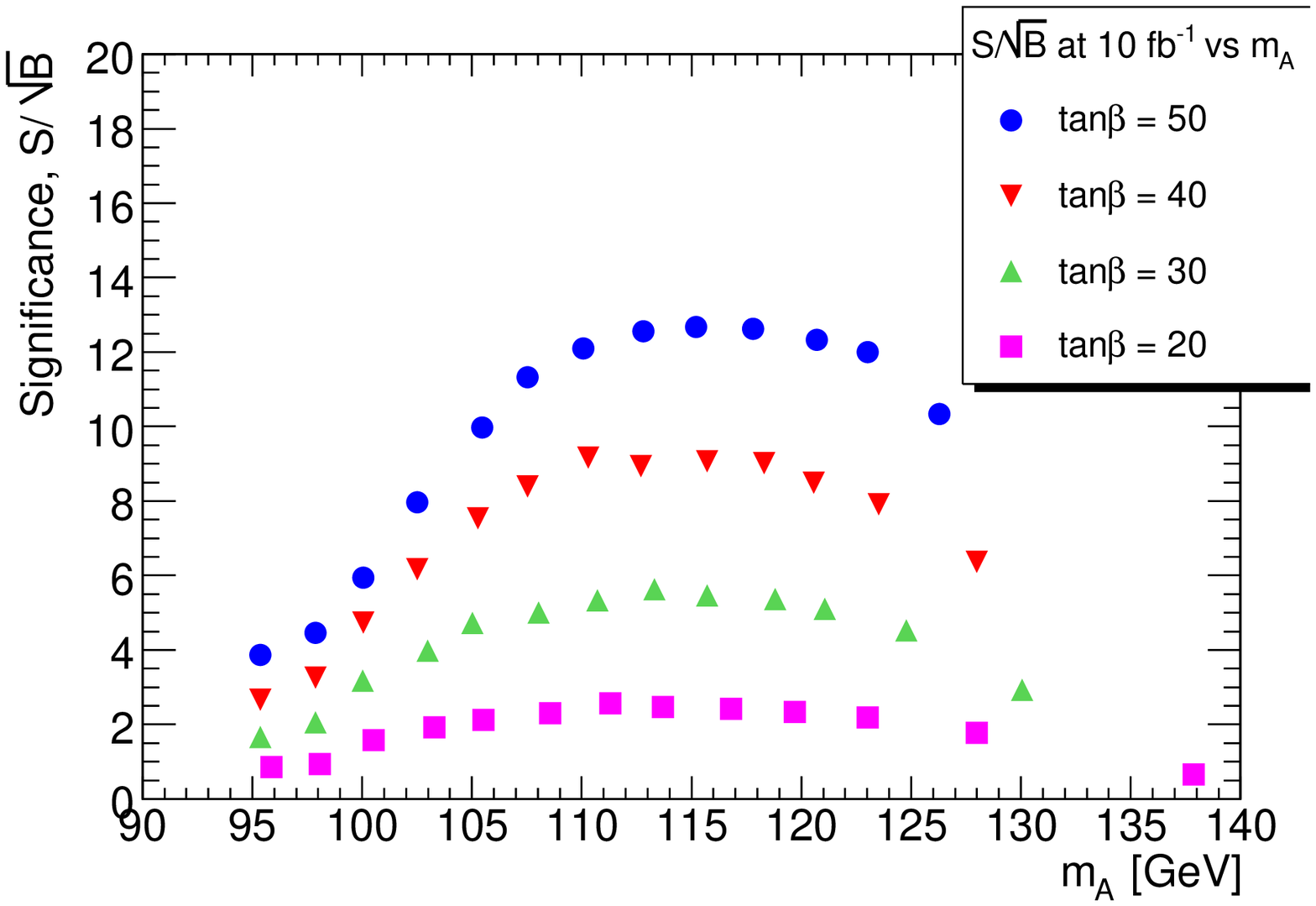} &
\hspace{-0mm}
\includegraphics*[width=0.5\textwidth]{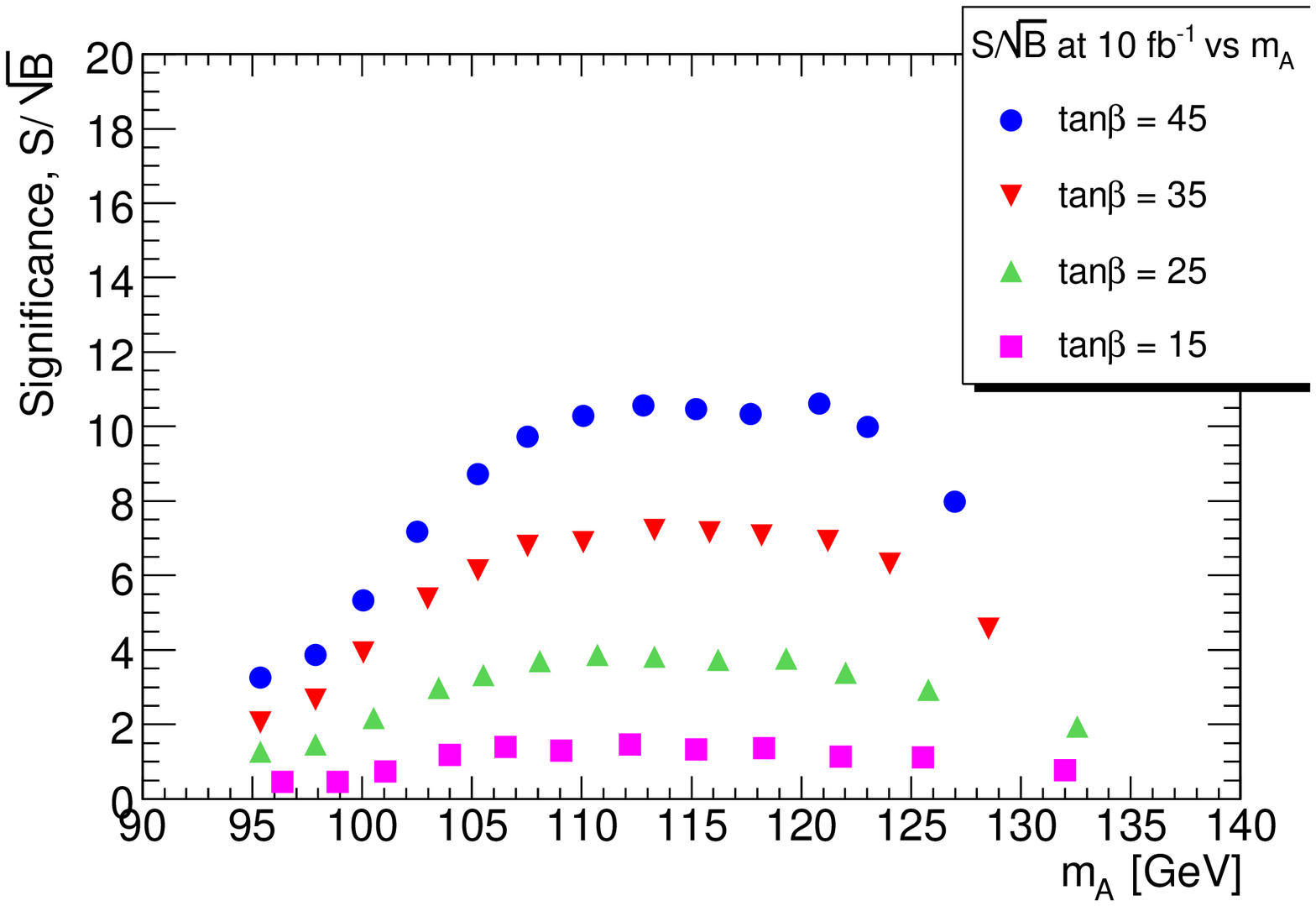}\\

\end{tabular}
    \caption[]{\label{fig:Significance}
    Search significance \SIGN\ for a h/A neutral Higgs boson, as a function of \MA\ up to the largest allowed value of 
    \Mh\  in three different data taking scenarios, \IL\ = 300, 30 and 10 \fbinv\ (S is the number 
    of h/A signal events, B is the number of background events).  
    On the left the results for \tanb\ = 50, 40, 30, 20, and on the right the results for \tanb\ = 45, 35, 25, 15.  
    The data are listed in Ref.\cite{Gentile07}.
}
\end{figure}
\begin{figure}
\begin{tabular}{cc}
\hspace{-7mm}
\includegraphics*[width=0.5\textwidth]{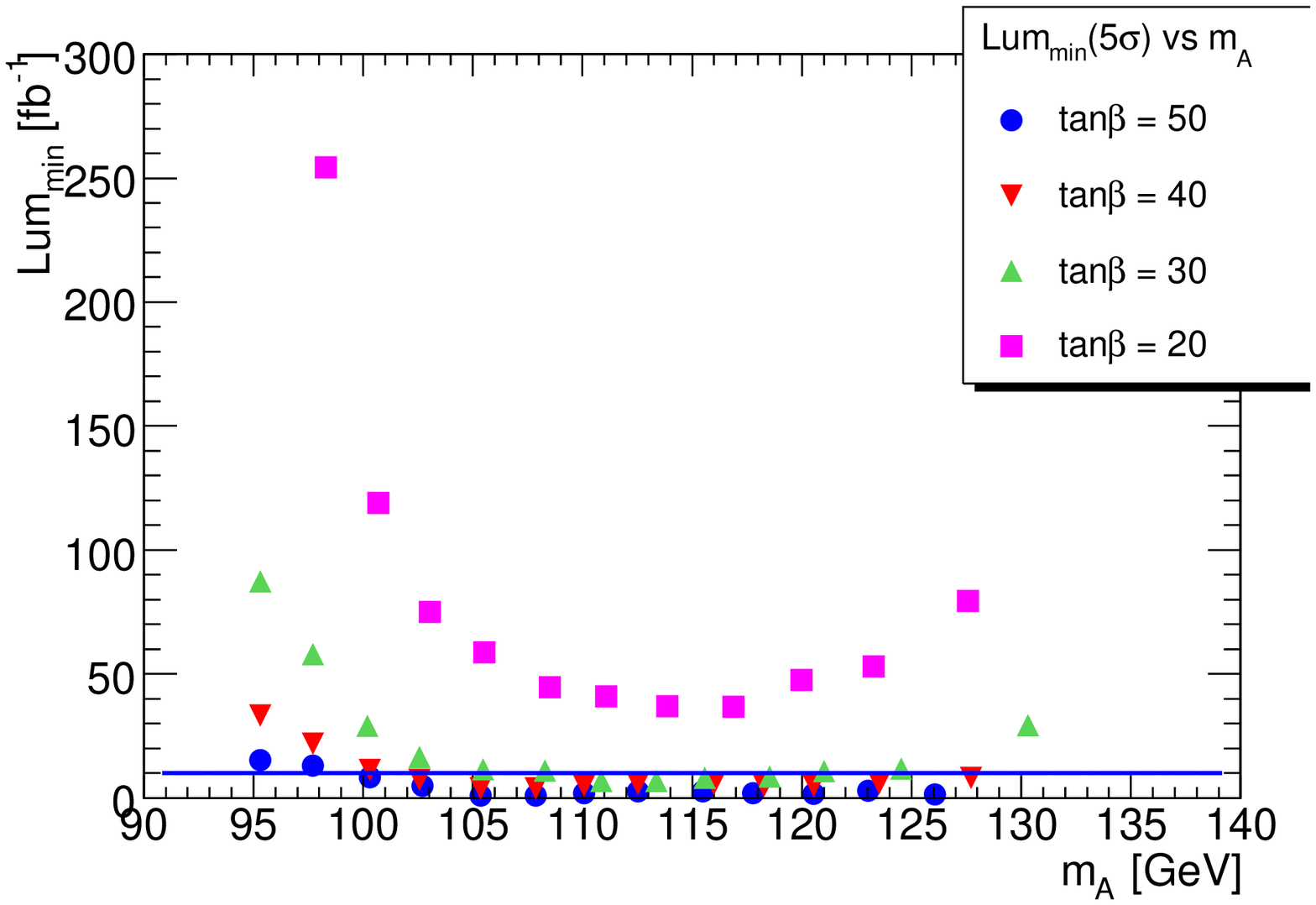} &
\hspace{-0mm}
\includegraphics*[width=0.5\textwidth]{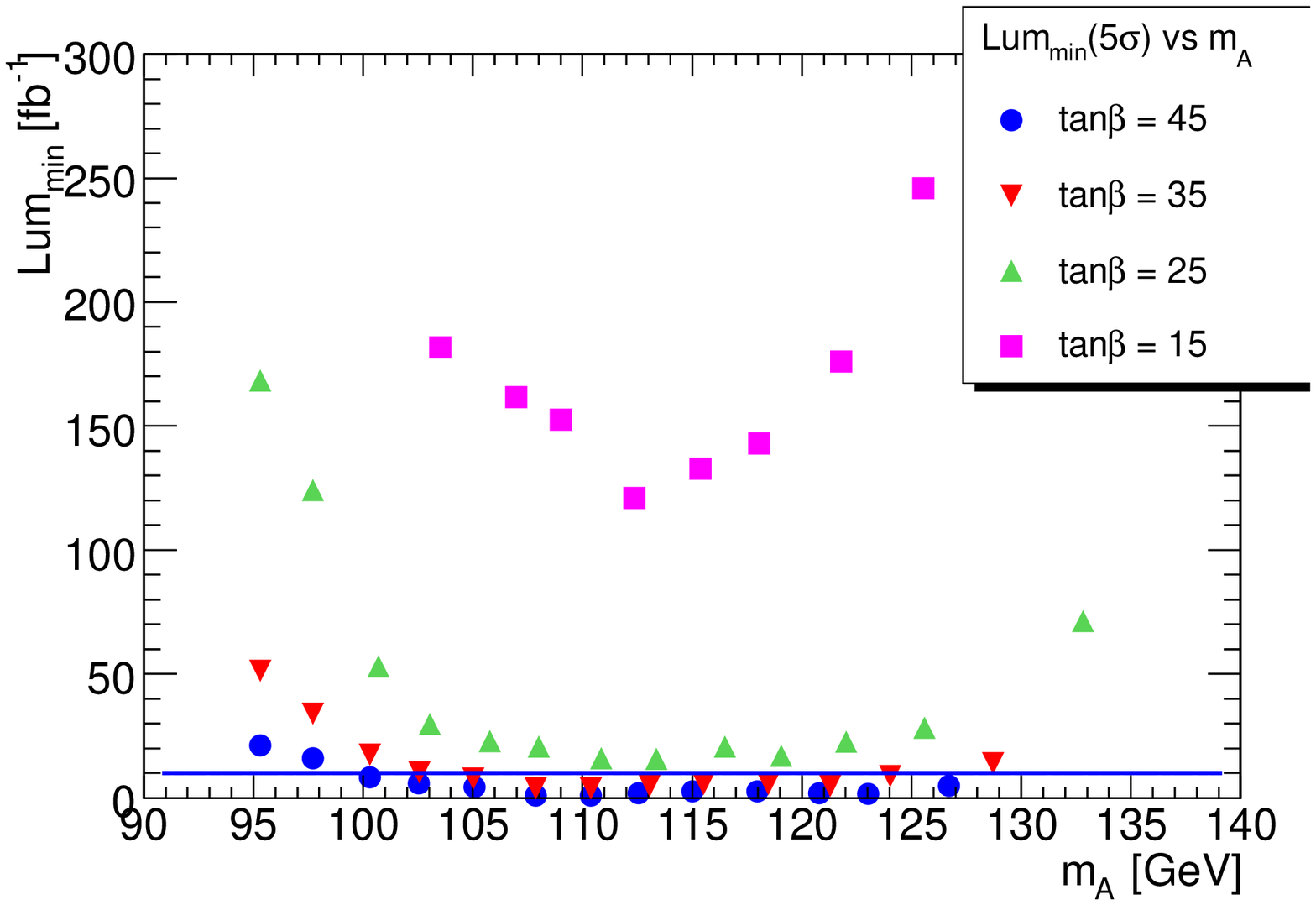} \\
\end{tabular}
    \caption[]{\label{fig:Luminosity_min}
     Minimum integrated luminosity \IL\ demanded for a 5 $\sigma$ discovery of the h/A neutral Higgs boson 
     as a function of \MA\ up to largest allowed value of \Mh . On the left the results for \tanb~ = 50, 40, 30, 20, 
     and  on the right the results for \tanb~ = 45, 35, 25, 15. The blue horizontal line is \IL\ = 10 \fbinv . 
     The data are listed in  Ref.\cite{Gentile07} .   
  }
\end{figure}
\par 
Discovery contours in the (\tanb , \MA ) plane are shown, in Fig.~\ref{fig:tanb_ma}, in different \IL\ scenarios  
for a significance of 5 (discovery, on the left) and of 3 (on the right). The latter can interpreted as the contour
region for an  early indication of a signal or, in case of negative search, for its exclusion.
\begin{figure}
\begin{center}
\begin{tabular}{cc}
\hspace{-7mm}
\includegraphics*[width=0.45\textwidth]{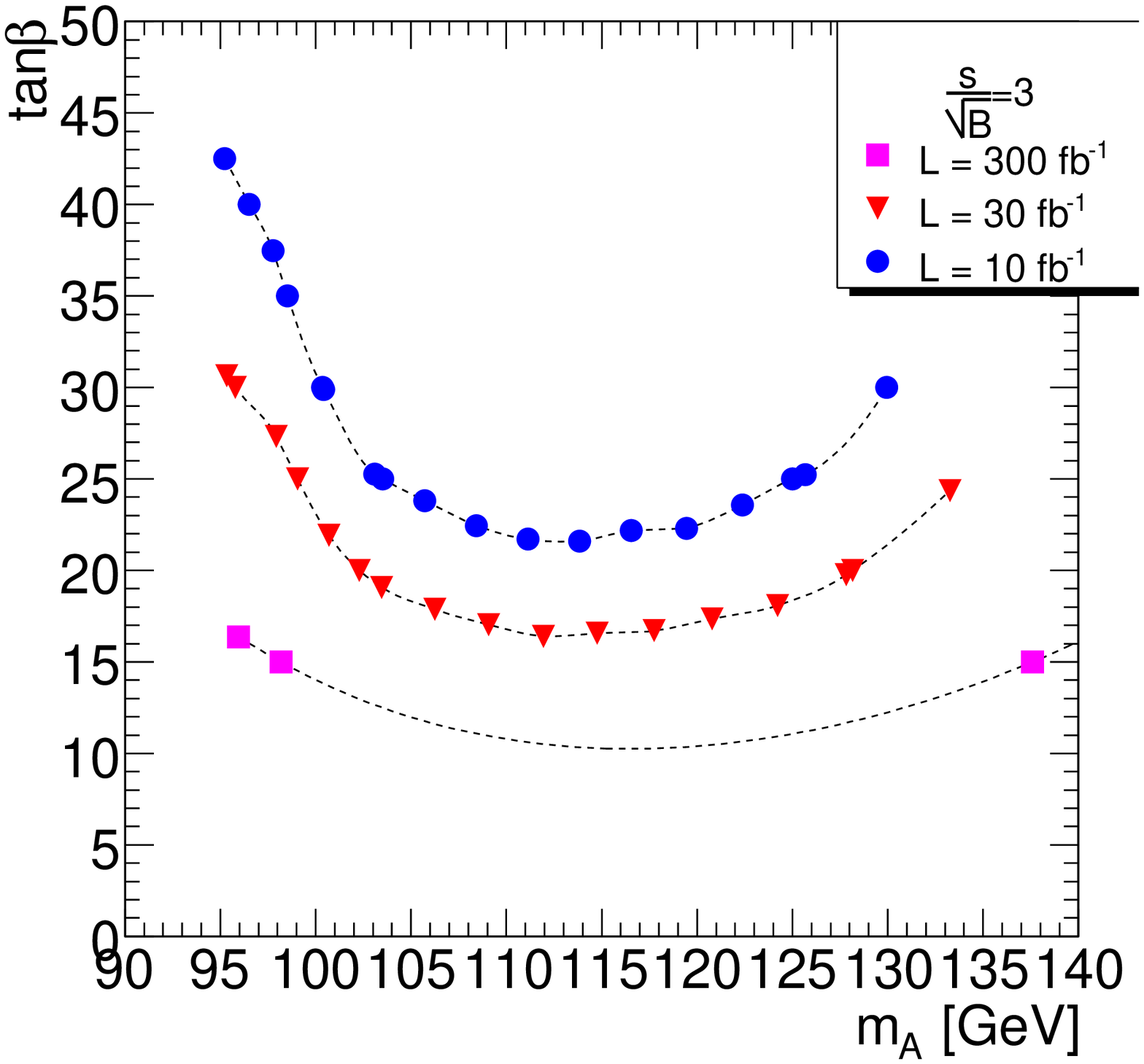} &
\hspace{-0mm}
\includegraphics*[width=0.45\textwidth]{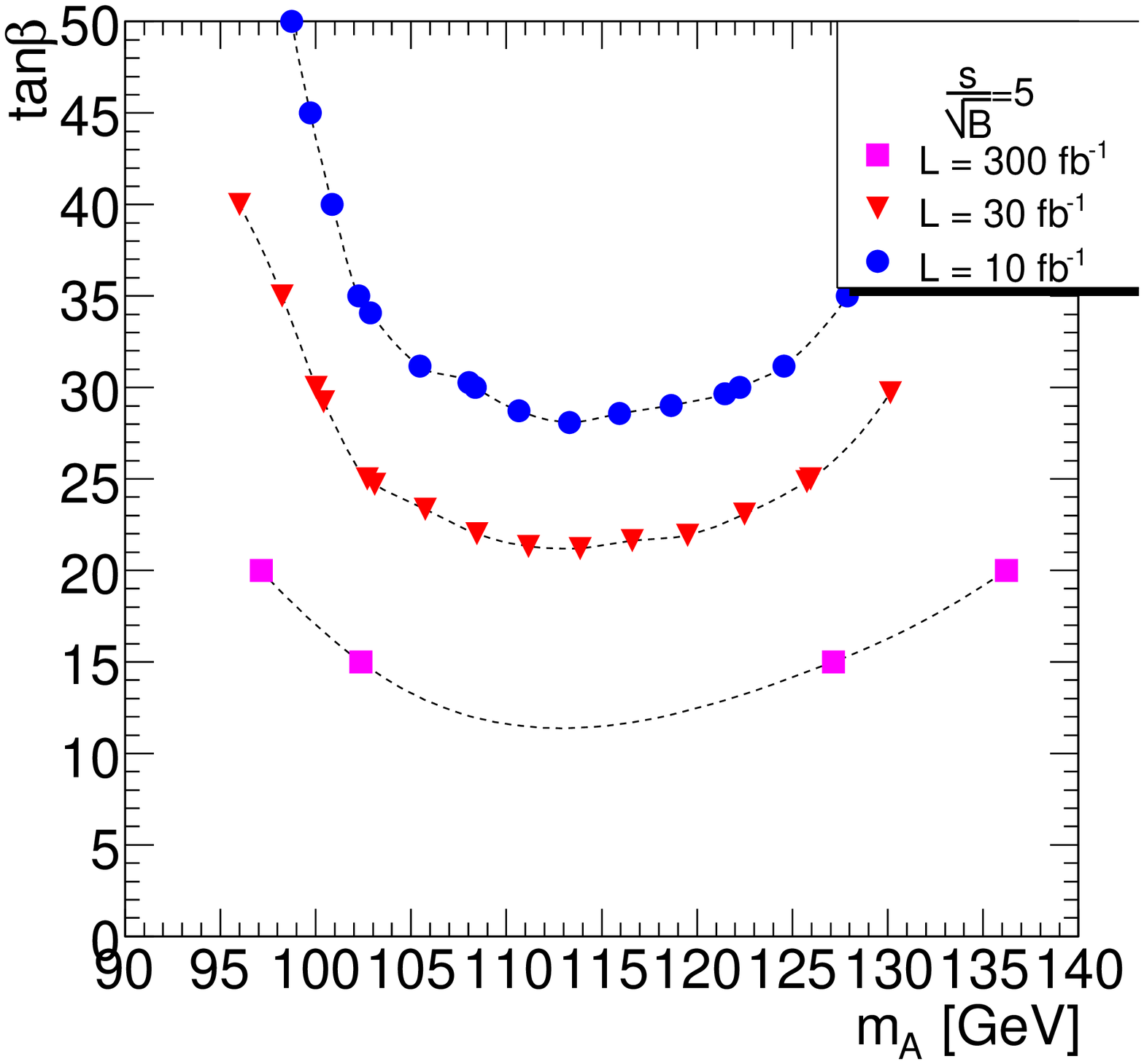} \\
\end{tabular}
    \caption[]{\label{fig:tanb_ma}

    Discovery potential for a neutral Higgs boson h/A  of mass \MA\ decaying to \mm , accompanied by two b-jets,
    in the $\Mh-{max}$ scenario (Sec.~\ref{sec:MSSM}), as a function of \MA : contours are drawn for   
    a search significance \SIGN~ = 5 (left) and  \SIGN~ = 3 (right), with an integrated luminosity of  \IL\ = 300 (top), 30
    (center)  and 10 (bottom) \fbinv .
 }
\end{center}
\end{figure}

\subsection{${\boldsymbol \BBH }$}\label{sec:Scan_H}

To increase the discovery potential  for a neutral Higgs boson discovery in the region up 139 \GeV\ a search for the neutral Higgs 
boson H has been performed in  the (\tanb , \MA ) plane inside the limits used for the h/A search (Sec.~\ref{sec:Scan_h}). This 
search required a separate analysis due to the higher mass region involved. 
\par
Due to the extremely low cross section of the H in this mass 
region, for a few masses  with a cross section times \mm\ branching ratio \SH $\cdot  \rm{Br}_{\mm}$ smaller than 0.01 \pb , common 
values of center and selection window were chosen, corresponding to the mean values in that mass region. This procedure doesn't 
affect any result due the narrow spread of values for the mass \MH\ and the natural width \GH , implying that the window  value is 
essentially dominated by the experimental resolution. 
\par
The discovery of the boson H demands a high integrated luminosity, $\IL\approx
300$ \fbinv , and it would be possible at masses corresponding to \Mh\ = 122.50 \GeV\ and 125 \GeV\ for values  of \tanb\ $\geq 30$.
At \tanb\ = 25, only a narrow range of \MH\ masses around 134 \GeV\ (corresponding to \Mh\ = 125 \GeV ) is accessible. Lower values 
of \tanb\ are excluded even at high masses.
\par 
In conclusion, a discovery of H boson is possible only in a few points of the  parameter space, for a value of \MA\ around the 
maximum value of \Mh , at the ultimate luminosity expected at the LHC.

\subsection{Combined search for ${\boldsymbol \BBA}$, \\ ${\boldsymbol \BBh}$, ${\boldsymbol \BBH~}$}\label{sec:Scan_ALL}

The results of  Sec.~\ref{sec:Scan_H} on the H boson search were  combined with  the results from the h/A search 
(Sec.~\ref{sec:Scan_h}). To this purpose, the H analysis was repeated with a window not overlapping with the h/A 
window, thus avoiding the double counting of background events. This work was  performed only for high \Mh\ values, where 
the cross section times the \mm\ branching ratio (Sec.~\ref{sec:Scan_H}) has values $\geq $ 0.01 \pb\ \cite{Gentile07}.  
\par
In Table \ref{tab:Significance_All} the significance of the exclusive search for the  h/A boson (\StrechA , \StrenhA , \SdiehA ) and 
the corresponding significance including the H search (\StrechAH , \StrenhAH~ and \SdiehAH ) at \IL\ = 300, 30, 10 \fbinv are shown.
\par 
The additional H search contributes to the MSSM Higgs sector discovery  for $\tanb  >  15$ at the ultimate luminosity.
Otherwise the contribution of this search is negligible, for a value of \MA\ around the maximum value of \Mh , as expected from the 
low cross section, and the large number of background events (being \MH\ close to \MZ ).

\begin{table}[!htb]
\begin{center}
\small
\begin{tabular}{|r||r|r||r|r|r||r|r|r|}
\hline
\bf {\tanb } & $\bf{\Mh^{\rm nom}}$ &\MH &\StrechA&\StrenhA&\SdiehA & \StrechAH & \StrenhAH&\SdiehAH \\
&[\SMeV]&[\MeV]&&&&&&\\
\hline
\hline

\bf{15}&\bf{122.50}  &  136180   &  3.86&    1.22&    0.70 & 5.09&  1.64 &    0.93  \\
\hline
\bf{15}&\bf{125.00} &   152610    &  1.51&    0.48&    0.28 &  2.27 &  0.73&    0.19  \\
\hline
\bf{20}&\bf{122.50}  &131870    &  9.76&    3.08&    1.78 & 11.69 & 3.77  & 2.14   \\
\hline
\bf{20}&\bf{125.00} & 139490   &   4.63&    1.47&    0.85 &  7.25 & 2.34  & 0.76 \\
\hline
\bf{25}&\bf{122.50}  & 130020   &  15.98&    5.05&    2.92 & 17.51  & 5.65  &3.2  \\
\hline
\bf{25}&\bf{125.00} & 134420    &  10.27&    3.25&    1.88 & 15.20  & 4.9   &2.78   \\
\hline
\bf{30}&\bf{122.50}  & 129090    &  25.13&    7.95&    4.59 & 27.01  & 8.71  & 4.94 \\
\hline
\bf{30}&\bf{125.00} & 131890     &  16.13&    5.10&    2.95 & 22.88  & 7.36   & 4.18\\
\hline
\bf{35}&\bf{122.50}  & 128560    &  34.71&    10.97&    6.34 & 35.29  & 11.38  & 6.45   \\
\hline
\bf{35}&\bf{125.00} & 130470   &  24.18&    7.65&    4.41 & 32.66  & 10.54   & 5.09  \\
\hline
\bf{40}&\bf{122.50}  & 128260   &  43.68&    13.81&    7.98 & 43.95   & 14.18 &8.03\\
\hline
\bf{40}&\bf{125.00} & 129610    &  34.02&    10.76&    6.21&  43.56 & 14.05 & 7.96     \\
\hline
\bf{45}&\bf{120.00}  & 127790  &  59.03&    18.67&    10.78&  54.16  &17.47  & 9.90  \\
\hline
\bf{45}&\bf{122.50}  & 128090    &  54.70&    17.30&    9.99 & 54.61  &17.62  & 9.98\\
\hline
\bf{45}&\bf{125.00} &  129090   &  43.54&    13.77&    7.95 & 53.90  & 17.38 & 9.85\\
\hline
\bf{50}&\bf{120.00}  & 127780   &  67.10&    21.22&    12.25 & 60.72  &19.59  &11.10  \\
\hline
\bf{50}&\bf{122.50}  & 128020    &  65.55&    20.73&    11.97&65.27   &21.10  & 11.93   \\
\hline
\bf{50}&\bf{125.00} & 128760   &  56.16&    17.76&    10.25   & 66.70  &21.52  &12.19  \\
\hline

\end{tabular}
\end{center}
\caption{Significance of  A/h and A/h/H searches.  The significance values for the two searches, \StrechA , 
\StrenhA , \SdiehA  and \StrechAH , \StrenhAH , \SdiehAH , at \IL\ = 300, 30, 10 \fbinv , respectively, are
given for a set of \tanb\ values and two values of the mass of the lightest neutral Higgs boson, 
$\Mh^{\rm nom}$. The mass \MH\ of the corresponding H Higgs boson is also noted.
 }
\label{tab:Significance_All}
\end{table}
\normalsize


\section{Conclusions}\label{sec:Concl}

The possibility of the discovery of the  MSSM h/A  bosons  in the region of high \tanb~( larger than 15) and mass close to 100 
\GeV\ has been investigated by exploiting the decay of the neutral h/A boson into two muons, \h $\ra \mm$ and \Aa  $\ra \mm$ , 
accompanied by two b-jets. This region is also accessible by charged MSSM Higgs boson \Hpm\ decays. 
\par
For this purpose an analysis using full detector simulation has been performed. Monte Carlo events have been generated for a 
center-of-mass energy  \RS= 14 \TeV\ through the ATHENA interface 
(v.9.0.4), while the ATLAS detector response has been simulated using the GEANT program through the ATHENA interface (v.10.0.1).
\par
The results described in this paper show a well defined possibility  for the discovery of a neutral Higgs boson in a  region 
traditionally difficult due to the presence of the \Zo\ resonance. 
\par
This is achieved thanks to the high resolution performance of the ATLAS detector, namely of the muon spectrometer and the inner 
detector, together with the high b-tagging capability. For completeness the search of H$\ra \mm$ has been explored in the same 
mass region.
\par 
The discovery of a neutral MSSM boson looks possible in a mass range of 100 to 120 GeV at $\tanb > 15 $, with an integrated 
luminosity \IL\ = 10 \fbinv , which corresponds to one year of data taking.  A warning should be made however because of
unaccounted uncertainties in background and other systematic errors that cannot be evaluated precisely at this time.  
\par  
With a view to perform this analysis on real data a method to subtract the main contributing background of Z boson decays 
Z \ra \mm\ has been suggested by us \cite{Gentile06}.  This method mainly relies on experimental data with limited Monte 
Carlo corrections. The procedure is based on the use of a control sample of Z bosons decaying to electrons, and  does not 
depend on complex theoretical calculations nor on their implementation in Monte Carlo. 

\section{Acknowledgments}\label{sec:Akno}
This work has been performed withins ATLAS Collaboration and we thank collaboration members for helpful discussions. We have 
made use of physics analysis framework and tools which are the result of collaboration-wide efforts.

\appendix
\section{Appendix:\\
subtraction of the ${\boldsymbol \BBZ }$ background} \label{sec:App_1}

Following the method outlined in Sec. \ref{sec:Prelim3} and based on the universality of the lepton coupling, the number 
of background events \BBZ, $N_{\Zo\ra \mm }$, is given by the number, $N_{\Zo\ra\ee}$, of \bb\Zo \ra\ee\ events collected 
in the same real data sample. In both cases the number of events is the one counted within a window of the two-lepton invariant 
mass distribution. The method has been fully proved with the simulation described in \cite{Gentile06}. 
\par   
The ratio $N_{\Zo\ra \mm}/N_{\Zo\ra \ee}$ has been shown to be a regular function of the dilepton invariant mass in the 
region of interest between 97.5 \GeV\ and 140 \GeV\ once possible sources of differences between \ee\ and \mm\ data  have 
been taken into account, and to be close to one. We like to recall here these differences and their impact on the above ratio. 
\par  
Clearly, depending on whether the final state contains electrons or muons, one has to provide for the differences in the 
detector response that is for the acceptance and resolution of the electromagnetic calorimeter and the muon spectrometer. 
\par 
Since our samples are $\bb \Zo \ra \mm$ or  $\bb \Zo \ra \ee$, in those events where a semileptonic b decay occurs with 
the same flavour as one of the \Zo\ decay products, more than one invariant mass combination is possible, namely three 
(only opposite charge lepton pairs are considered for invariant mass combinations). Often, the value of the invariant 
mass associated to this combination of one lepton originating from \Zo\ and another one from b decay sits far from \MZ\
(and thus has no effect in the interested region). However the higher efficiency in muon detection increases the number of 
{\it fake} combinations in the muon sample.  
\par
Less obvious are the corrections required by the different {\it inner bremsstrahlung} in the two samples. The {\it inner 
bremsstrahlung } (IB) is the emission of photons, \gbrem, near the Z\ee\ or Z\mm\ vertex. The presence of such photons changes the 
kinematic configuration of the decay, in particular the lepton 4-momenta. The impact of this effect on the Higgs mass 
resolution with the ATLAS detector has been studied in Ref.\cite{Linossier95}. Here we discuss its impact on the number of 
detected events as a function of the dilepton reconstructed invariant mass in the region of our search. 
\par
At the generator level in the electron (muon) sample 85.2\%  (91.6\%) of the events are without IB photons, 13.7\% (8\%) have 
1 \gbrem, 0.94\% (0.2\%) have 2 \gbrem. The corresponding average transverse momentum of the radiated photons, without any cut 
except those mentioned in Sec. \ref{sec:MC}, is close in the two samples, $<\PT^{\gbrem}>  = 16.18~\GeV$ for the \ee\ sample 
and $<\PT^{\gbrem}> = 16.43~\GeV$ for the \mm\ sample.  These IB photons are mainly contained in a cone with an opening $\DR = 
0.15$ around a muon track and $\DR = 0.25$ around an electron track. By summing up the IB photon 4-momentum to the close lepton
4-momentum one obtains that the ratio  $N_{\Zo\ra \mm}/N_{\Zo\ra \ee}$ stays the same within $\pm 12\%$ for dilepton invariant
masses between 100 and  140 \GeV. It would otherwise show a spread of $\approx \pm 30\%$. 
\par 
At the detector level 
muons, electrons  and photons involve  different detectors (muon spectrometer and electromagnetic calorimeter) implying different 
momentum resolution, angular acceptance and efficiency. In this analysis we used identification criteria as follows.
\par
A muon is reconstructed as a track, in both the inner detector and the muon spectrometer (so named combined reconstruction), and
enters the analysis if $\PT >10~\GeV$ and $|\eta|<2.5$. The average momentum reconstructed is  \hbox{ $<\PT> = 37.83~\GeV$}. The 
reconstruction efficiency is higher for muons than for electrons. 
\par 
An electron is reconstructed as a cluster in the electromagnetic calorimeter associated with a track inside $\Delta \eta =0.025$ and 
$\Delta \phi =0.05$, and enters the analysis if  $\PT >10~\GeV$ and $|\eta|< 2.5$. The best performance is achieved when the energy 
is measured in the electromagnetic calorimeter (with a cluster size of $5\times 5$ cells) and the angles ($\theta,\phi$) in the 
tracker. The average momentum reconstructed is \hbox{ $<\PT> = 42.61~\GeV$}. 
\par
A photon is  reconstructed as a cluster in the electromagnetic calorimeter not associated to any track inside $\Delta \eta =0.025$ 
and $\Delta \phi =0.05$, and enters the analysis if $E_{{\rm T}} >10~\GeV$ and $|\eta|<2.5$. A cluster size of 5$\times$5 cells is 
used to measure the energy. The value chosen for the $E_{{\rm T}}$ threshold ensures a good photon reconstruction efficiency.
\par
In this analysis however a photon's 4-momentum is summed with that of the lepton if the photon is reconstructed  inside a cone $\DR 
< 0.15$ around an electron (\ee\ sample) or $\DR < 0.25$ around a muon (\mm\ sample). As a result, although the number of photons of 
any origin is more important in the \ee\ than in the \mm\ sample, the number of photons reconstructed separately is equal ($\sim 1\%$) 
in the two samples. The effect of summing the photon and lepton 4-momenta shows up in the distribution of the ratio 
$N_{\mm}/N_{\ee}$ as a function of the dilepton invariant mass when only events containing an IB photon at the generator level 
are considered. It consists in a shift from the low-mass tail towards the \Zo\ mass.
\par
Based on the whole simulation we could conclude \cite{Gentile06} that the ratio $N_{\mm}/N_{\ee}$ is $\approx 1.2$ and does not
depend, in the region of our search and within the simulation statistics, on the two-lepton invariant mass.